\documentclass[prb,
 superscriptaddress, twocolumn,nofootinbib,
aps,showpacs]{revtex4-1}

\usepackage{graphicx}
\usepackage{hyperref}
\usepackage{amsmath}
\usepackage{url}
\usepackage[dvipsnames]{xcolor}
\usepackage{verbatim}
\usepackage{gensymb}
\usepackage{subfig}

\def\beq {\begin{equation}}
\def\eeq {\end{equation}}
\def\w {\omega}
\def\bfq {\mathbf{q}}

\def\bfG {\mathbf{G}}
\def\bfk {\mathbf{k}}
\def\bfr {\mathbf{r}}

\newcommand{\bra}[1]{\langle #1|}
\newcommand{\ket}[1]{|#1\rangle}
\newcommand{\expval}[1]{\langle #1 \rangle}

\newcommand{\rv}{\mathbf{r}}
\newcommand{\kv}{\mathbf{k}}
\newcommand{\qv}{\mathbf{q}}
\newcommand{\qvt}{\mathbf{\tilde{q}}}
\newcommand{\gv}{\mathbf{G}}
\newcommand{\rhot}{\tilde{\rho}}

\date{\today}
\begin{document}

\newcommand{\lsi}{Laboratoire des Solides Irradi\'es, \'Ecole Polytechnique, CNRS, CEA/DRF/IRAMIS, Institut Polytechnique de Paris, F-91128 Palaiseau, France}
\newcommand{\etsf}{European Theoretical Spectroscopy Facility (ETSF)}
\newcommand{\soleil}{Synchrotron SOLEIL, L'Orme des Merisiers, Saint-Aubin, BP 48, F-91192 Gif-sur-Yvette, France}

\author{Arnaud Lorin }
\affiliation{\lsi}
\affiliation{\etsf}

\author{Matteo Gatti}
\affiliation{\lsi}
\affiliation{\etsf}
\affiliation{\soleil}

\author{Lucia Reining }
\affiliation{\lsi}
\affiliation{\etsf}

\author{Francesco Sottile}
\affiliation{\lsi}
\affiliation{\etsf}

\title{First-principles study of excitons in the optical spectra of silver chloride}

\begin{abstract}
 Silver chloride is a material that has been investigated and used for many decades. Of particular interest are its optical properties, but only few fundamental theoretical studies exist. We present first-principles results for the optical properties of AgCl, obtained using time-dependent density functional theory and many-body perturbation theory. We show that optical properties exhibit strong excitonic effects, which are correctly captured only by solving the Bethe-Salpeter equation starting from  quasiparticle self-consistent GW results. Numerical simulations are made feasible by using a model screening for the electron-hole interaction in a way that avoids the calculation of the static dielectric constant. A thorough analysis permits us to discuss localization in bright and dark excitons of silver chloride.
\end{abstract}

\maketitle

\section{Introduction}

Silver chloride is a versatile material, long known for a large variety of applications. For instance, it is a reference 
electrode for electrochemical measurements\cite{Bates1978}, and in its nanostructured phase
it has remarkable antimicrobial properties \cite{Kang2016,Trinh2015,Adams1999}. Moreover, it has been recently shown that silver clusters at AgCl surfaces form an efficient photocatalytic system 
\cite{Schuerch2002,Wang2008,Lou2011,Zhang2013,Cai2014}. 
The largest range of applications of AgCl is  related to its optical properties: it is responsible for several shades in stained glass \cite{Jembrih2002}, and
it is widely used as photochromic material in photosensitive glasses  \cite{Armistead1964}. In particular, AgCl is a  
crucial ingredient in photographic paper  
to produce the latent image\cite{Eachus1999,photographic}. 
It was already the key component in the first color photography in history realised by E. Becquerel in 1848 \cite{Becquerel1848b}: a recent study has shown  that the colors in Becquerel's photochromatic images  were due to silver nanoparticles in a silver chloride matrix \cite{deSeauve2020,deSeauve2020b,deSeauve2020c}. 

In spite of the importance of its optical properties, 
experimental results, including absorption,  reflectivity and luminescence, are relatively old\cite{Mott1948,Seitz1951,Moser1956,Aline1957,Tutihasi1957,Brown1961,Brown1962,Carrera1971,Hartman:61,White:68,Bauer:74,Yanagihara1983} (see Refs. \cite{vonderOsten1984,Ueta1986} for more recent reviews). 
Moreover, the same photochromic properties of AgCl that make it so appealing for applications also hamper spectroscopy experiments: 
its electronic and optical properties can be changed significantly by irradation with light \cite{Victor}, thus affecting the reliability of the measured spectra. 
On the other side, theoretical simulations are a valuable tool to provide  a solid benchmark and remove possible ambiguities from experimental results.
Several first-principles studies\cite{Ma2012,Benmessabih2007,Amrani2007,Zaoui2005,Okoye2002,deBoer1999,Vogel1998} within density functional theory\cite{HK,KS} (DFT) have focused on ground-state properties and the Kohn-Sham electronic structure. However, these methods cannot access the band gap, a fundamental ingredient for the optical properties. Only recently, band structure calculations using the GW approximation  \cite{Hedin1965} (GWA) within many-body perturbation theory 
 \cite{Fetter1971} (MBPT) have yielded more reliable numbers for the photoemission gaps \cite{vanSetten2017,Gao2018,Zhang2019}. Instead, to the best of our knowledge, MBPT studies for the optical properties of silver chloride are still missing.

The present work aims at bridging this gap: we have conducted state-of-the-art electronic structure calculations to investigate the optical properties of bulk 
AgCl. Indeed, besides  its strong interest for a wide range of applications, AgCl is challenging from the theoretical point of view: Ag $4d$ states are strongly hybridized with Cl $3p$ and  have  a direct impact on the band gap. Their strong localization requires a high cutoff in plane wave calculations, and it moreover poses problems for simple density functionals such as the local density approximation (LDA) or the generalised gradient approximation (GGA).
Simple  models \cite{Bassani1965,Wang1976} are not reliable because of its peculiar band structure
and estimates of excitonic effects based on the Wannier model (see e.g. Refs. \cite{vonderOsten1984,Ueta1986}) should be examined with great care: advanced first-principles approaches are needed in order to get reliable insight.

The questions that we will address in the present work are the following: 
\textit{ Which level of theory is needed for a proper description of optical properties of AgCl, including questions related to pseudopotentials, self-consistency, and excitonic effects? Can we simplify the calculations, in spite of the complexity of the material? How strong are excitonic effects due to the electron-hole interaction? } 
Finally, \textit{The nature of excitons} in AgCl is a crucial question with a direct impact on all the applications that involve the optoelectronic properties of AgCl.

After this introduction, Sec. \ref{sec:methodology} summarizes the approaches used to access ground- and excited-state  properties. Results for the 
band structures from DFT and MBPT, as well as first results for the optical properties, are given in Sec. \ref{sec:results}. In Sec. \ref{sec:model}, we propose a way to efficiently use a model screening of the electron-hole interaction, and we show that this allows us to overcome the computational limitations and obtain reliable optical spectra. Finally, Sec. \ref{sec:excitons} is dedicated to the discussion of excitonic effects in the optical properties of AgCl, and conclusions are drawn in Sec. \ref{sec:conclusions}.

\section{Methodology}
\label{sec:methodology}

Starting from a DFT ground-state calculation, there are two possible routes to determine optical spectra\cite{Onida2002}: first, to extend DFT to time-dependent DFT\cite{Runge1984,Ullrich2012} (TDDFT), and second, to move to MBPT\cite{Fetter1971,the-book}, where the one-body Green's function is determined from a Dyson equation with a given approximation for the self-energy, and subsequently optical properties are derived from the solution of a Bethe-Salpeter equation\cite{Bethe1951,Strinati1988} (BSE) for the two-body correlation function. When simple approximation for the exchange-correlation (xc) functionals are sufficient, TDDFT is computationally more efficient than MBPT. It is therefore interesting to compare the results of the two approaches, and we will detail both routes in the following.

\subsection{Pseudopotentials}

All our calculations are carried out in a plane wave basis:
wavefunctions are written as linear combination of plane waves $\psi_{n\kv}(\rv)=\frac{1}{\sqrt{\Omega}}\sum_{\gv=0}^{\gv_{max}}u_{n\kv}(\gv) e^{i(\bfk+\gv)\cdot\rv}$, 
where $\gv$ are reciprocal-lattice vectors and the wavefunctions are normalized by the crystal volume $\Omega$. 

Since the presence of core electrons would require an unaffordably large plane-wave cutoff  $\gv_{max}$, we explicitly take into account only the valence electrons and use pseudopotentials to represent the cores. Our pseudopotentials are of 
Troullier-Martins \cite{Troullier1991} type for both species, silver and chlorine. 
We have generated the pseudopotential using the FHI98PP package\cite{FUCHS199967}. We used an  LDA xc functional in the Perdew-Wang parametrization \cite{Perdew1992} with scalar relativistic corrections.

The ground-state configuration of Ag is 
\[ 1s^2 2s^2 2p^6 3s^2 3p^6 4s^2 3d^{10} 4p^6 5s^1 4d^{10}  \mbox{ or  } 
[Kr]5s^1 4d^{10}.\] However, pseudopotentials are usually created in a slightly ionized state of the atom \cite{Bachelet1982}. Moreover, it is generally established that spectroscopy calculations require valence shells to be complete, because of the strong spatial overlap between electrons in the same shell and the consequent strong exchange effects \cite{Rohlfing1995,Marini2001,Bruneval2007}. Here we use the atomic  configuration
$$ 1s^2 2s^2 2p^6 3s^2 3p^6 4s^2 3d^{10} 4p^6 4d^{10} $$ to create the pseudopotential of silver, with the 4$spd$-shell 
 $ 4s^2 4p^6 4d^{10}$ in the valence. We use $s$ as the local reference component to represent all higher angular momenta. Cutoff radii were set to 0.9 a.u. for the $s$, 2.3 a.u. for the $p$, and 1.5 a.u. for the $d$ component, guaranteeing satisfactory logarithmic derivatives and excitation energies.
 
 Chlorine does not present the same difficulty, and we have created the pseudopotential using an atomic configuration of $1s^2 2s^2 2p^6$ for the core and $ 3s^2 3p^{4.5} 3d^{0.5} $ for the valence electrons. 
Cutoff radius were $1.6$ a.u. for $s$ and $p$ components and $1.8$ a.u. for the $d$ component.

\subsection{Ground-state density functional theory and the Kohn-Sham electronic structure}

For the ground-state calculations we use DFT \cite{HK} in the formulation of Kohn and Sham (KS) \cite{KS}, i.e. we solve the single-particle KS equations  
\begin{multline}
 \left[ -\frac{1}{2} \nabla^2 + v_{\rm ext}(\rv) + v_H([n],\rv) + v_{\rm xc}([n],\rv) \right] \psi_{n\kv}(\rv)  
 \\ = \varepsilon_{n\kv} \psi_{n\kv}(\rv),  \label{dft_ks}
\end{multline}
where the external potential $v_{\rm ext}(\rv)$ is the pseudopotential due to the ions\footnote{To be precise, since we use pseudopotentials the external potential is non-local, $v_{\rm ext}(\rv.\rv')$.},  $v_H(\rv)$ is the Hartree potential, and  we use the LDA \cite{Ceperley1980,Goedecker1996} for the xc potential $v_{\rm xc}([n],\rv)$.
$n,\kv$ label band and crystal momentum within the first Brillouin zone, and $\varepsilon_{n\kv}$ and  $\psi_{n\kv}(\rv)$ are, respectively, eigenvalues and eigenvectors of the KS Hamiltonian. 

Calculations are done using the Abinit package \cite{Gonze2005}.
Converged results for AgCl were obtained by using  Monkhorst-Pack \cite{Monkhorst1976} $8\times 8 \times 8$ grids shifted along 4 directions  and a kinetic 
energy cutoff $E_{cut} = G^2_{max}/2 = 150$ Hartree.
This high cutoff is needed because of the strongly localized semi-core $4s$ and $4p$ states of silver. Since we are dealing with a non-magnetic material, here and in the following we omit spin, which will only lead to prefactors. Atomic units are used in this paper.

\subsection{Optical absorption with time-dependent density functional theory}

The optical properties of a systems are linked to the inverse dielectric function $\epsilon^{-1} = 1 + v_c\chi$, where $v_c$ is the Coulomb potential and $\chi$ is the linear density-density response function. This quantity can be accessed, in principle exactly, using TDDFT in linear response \cite{Petersilka1996}. First, one has to build the non-interacting polarizability, which is in  frequency and reciprocal space a function of frequency $\omega$ and of momentum $\qv$ in the first Brillouin zone, and a matrix in reciprocal-lattice vectors,
\begin{widetext}
\begin{equation}
\chi^0_{\gv\gv'}(\qv,\omega) = \frac{2}{N_k\Omega_0}\sum_{n_1n_2\kv} \left(f_{n_1}-f_{n_2} \right) 
 \frac{\rhot_{n_1n_2\kv}(\qv,\gv)\rhot^*_{n_1n_2\kv}(\qv,\gv')}{\omega - (\varepsilon_{n_2\kv}-\varepsilon_{n_1\kv-\qv})+i\eta }.  
 \label{chi0}
\end{equation}
\end{widetext}
Here 
$f_{n_i}$ are occupation numbers. 
$N_k$ is the 
number of $\bfk$ points in the first Brillouin zone and $\Omega_0$ the volume of the unit cell. 
The positive infinitesimal $\eta$ ensures causality; a non-vanishing value gives rise to a Lorentzian broadening. The factor $2$ stems from the sum over spins. The matrix elements $\rhot_{n_1n_2\kv}(\qv,\gv)= \int \psi^*_{n_1\kv-\qv}(\rv) e^{-i(\qv+\gv)\cdot\rv}\psi_{n_2\kv}(\rv) d\rv$
give the oscillator strengths. 
For optical properties, the wavevector is very small compared to the crystal, and we take the limit
$\qv\to 0$ (where the transverse and longitudinal dielectric functions coincide \cite{DelSole1993}).

The full density-density response function is then obtained from  the Dyson-like linear-response equation\cite{Petersilka1996}
\begin{equation}
\chi = \chi^0 + \chi^0 \left( v_c+ f_{\rm xc} \right) \chi,
\label{dysonchi}
\end{equation} 
where in reciprocal space all quantities are functions of $\qv$ and (besides $v_c$) of $\omega$, and matrices in $\gv,\gv'$. 
The exchange-correlation kernel $f_{\rm xc}(\rv,\rv',t,t')\equiv\delta v_{\rm xc}(\rv,t)/\delta n(\rv',t')$ is the functional derivative of the xc potential with respect to the density $n$. It depends on two space (or reciprocal space) arguments and on the time difference $t-t'$ (or frequency $\omega$). Its exact expression is unknown.
Two extensively used approximations are the random-phase approximation (RPA) $f_{xc}\approx 0$, and the adiabatic local density approximation (ALDA), where
$f_{\rm xc}(\rv,\rv',t,t')\approx\delta(\rv-\rv')\delta(t-t')d v_{\rm xc}^{\text{\tiny LDA}}(\rv,t,n(\bfr,t))/d n(\rv,t)$.  

From the density-density response function we evaluate the inverse dielectric function
\begin{equation}
    \epsilon^{-1}_{\gv,\gv'}(\qv,\omega) = \delta_{\gv,\gv'} +
    \frac{4\pi}{|\qv+\gv|^2}\chi_{\gv,\gv'}(\qv,\omega).
\end{equation}
The macroscopic dielectric function \cite{Adler1962,Wiser1963} is then 
\begin{equation}
\epsilon_M(\omega) = \lim_{\qv\to 0}\frac{1}{\epsilon^{-1}_{\gv=\gv'=0}(\qv,\omega)}.
\label{epsm}
\end{equation}
From the macroscopic dielectric function we derive optical properties: optical absorption is related to the imaginary part,  ${\rm Im}\,\epsilon_M$, and the extinction coefficient is given by 
$\kappa ={\rm Im }\,\sqrt{\epsilon_M}$. Eq. \eqref{epsm} takes into account crystal local field effects, because 
the dielectric matrix is inverted before the macroscopic average $\gv=\gv'=0$ is taken. In the RPA and when local fields are neglected, the macroscopic dielectric function becomes
\begin{eqnarray}
\epsilon_M(\w) &=&  1 - \lim_{\qv\to 0}\frac{8\pi}{N_k\Omega_0 q^2} \sum_{vc\kv} \left[\frac{\big| \tilde{\rho}_{vc\kv}(\qv) \Big|^2 }{\w -(\varepsilon_{c\kv} - \varepsilon_{v\kv})+ i\eta}\right. \nonumber\\
&-& \left.\frac{\big| \tilde{\rho}_{cv\kv}(\qv) \Big|^2 }{\w +(\varepsilon_{c\kv} - \varepsilon_{v\kv})+ i\eta}\right].
\label{epsnlf}
\end{eqnarray}

The linear response TDDFT calculations were carried out using the DP code \cite{DPcode}. Convergence for both absorption and the extinction coefficient over a frequency range of 0 to 10 eV was reached using 2048 shifted 
$\bfk$ points \cite{Benedict1998}, 965 plane waves for the wave functions, 59 $\gv$ vectors for the polarizability matrix including crystal local field effects, 
13  occupied bands, and 20 unoccupied bands. 

\subsection{Band structure with the GW approximation}
\label{sec:gw-technical}

KS eigenvalues cannot be interpreted as electron removal and addition energies, but they often give a good overview of the band structure and constitute a convenient starting point for further calculations. In  order to obtain a more meaningful band structure, we add quasiparticle corrections using the Green's function formalism.
In the quasiparticle approximation, addition and removal energies are obtained from a modified one-particle equation
\cite{Hedin1965,the-book}
\begin{multline}
 \left[ -\frac{1}{2} \nabla^2 + v^{\rm ext}(\rv) + v_H([n],\rv) \right] \phi_{n\kv}(\rv)  \\ + \int d\rv' \Sigma_{\rm xc}(\rv,\rv',E_{n\kv}) \phi_{n\kv}(\rv') 
  = E_{n\kv} \phi_{n\kv}(\rv),  \label{qp}
\end{multline}
where the self-energy $\Sigma_{\rm xc}$ plays the role of an effective non-local and energy-dependent potential. The generalized eigenvalues of Eq. \eqref{qp} can be interpreted as addition and removal energies, and are used to build the theoretical band structure. The quasiparticle wavefunctions $\phi_{n\kv}$
are also in principle different from the KS ones, which changes in particular the density. 
A widely used approximation for the self-energy is Hedin's GWA\cite{Hedin1965}.
In this approximation, $\Sigma_{\rm xc}=iGW$ is the product of the one-body Green's function $G$ and the screened Coulomb interaction $W=\epsilon^{-1}v_c$.

Usually a quasiparticle approximation is made for the Green's function $G$ that is needed to build the GW self-energy.
With the quasiparticle spectral weight normalized to 1, the Green's function can then be written as
\begin{equation}
G(\rv,\rv',\omega)\approx G^0(\rv,\rv',\omega)=\sum_{n\kv}\frac{\phi^*_{n\kv}(\rv)\phi_{n\kv}(\rv')}{\omega - E_{n\kv}+i\eta}.
\label{eq:qpapp}
\end{equation}
Still, its knowledge requires in principle  the solution of Eq. (\ref{qp}), which makes   the problem self-consistent. Many calculations replace the quasiparticle wavefunctions and eigenvalues by KS ones. Moreover, they evaluate the screened Coulomb interaction $W$ using the RPA for $\epsilon^{-1}$ following 
Eqs. \eqref{chi0} and \eqref{dysonchi}. These two approximations define the G$^0$W$^0$ approach\cite{Hybertsen1986,Godby1988,Aryasetiawan1998,Aulbur1999}. 

A further simplification is obtained by 
evaluating the quasiparticle eigenvalues perturbatively with respect to the KS ones, and by making use of the fact that the self-energy is approximately linear around the quasiparticle energy. Using Eq. \eqref{qp} and Eq. \eqref{dft_ks}, this yields
\begin{equation}
E_{n\kv} = \varepsilon_{n\kv} + Z_{n\kv} \big[ \expval{\Sigma_{\rm xc}(\varepsilon_{n\kv})} - \expval{v_{\rm xc}} \big],
\label{eigen}
\end{equation}
with the quasiparticle renormalization factor $Z_{n\kv}=\left[1 - \expval{\left.\frac{\partial \Sigma_{\rm xc}(\w)}{\partial\w}\right|_{\varepsilon_{n\kv}}} \right]^{-1}$.
Here, expectation values are taken with the KS  wavefunctions $\psi_{n\kv}$.

The G$^0$W$^0$ approach based on KS calculations with approximate functionals such as the LDA or GGA has met broad success for many  
materials\cite{the-book,Aryasetiawan1998,Aulbur1999,Bechstedt2014}, but it encounters problems when it comes to materials with localized electrons \cite{the-book,Faleev2004,Schilfgaarde2006,Bruneval2006,Bruneval2014,Bechstedt2014}. These are often transition metal oxides and other correlated materials where $d$ or $f$ electrons are important, but as we will see, the problem also concerns AgCl, because of the hybridisation between  Ag $4d$ and Cl $3p$ electrons. 
These materials require better starting eigenvalues and wavefunctions, or self-consistency. A prominent self-consistent approach is
quasiparticle self-consistent GW \cite{Faleev2004,Schilfgaarde2006} (QSGW). In this approach,
 Eq. \eqref{qp} is approximated by an effective Schr\"odinger equation with a static hamiltonian, and the resulting eigenvalues and eigenfunctions are used to build a new quasiparticle Green's function and screened Coulomb interaction. The procedure can be iterated to self-consistency and often improves over G$^0$W$^0$ results \cite{the-book,Faleev2004,Schilfgaarde2006,Bruneval2006,Bruneval2014,Bechstedt2014}.

We have performed band-structure calculations with both  G$^0$W$^0$ and QSGW, using the Abinit package \cite{Gonze2005} in a plane-wave basis and with a 4 times shifted $(4\times4\times4)$ grid to sample the Brillouin zone. 
For G$^0$W$^0$ calculations,
$W$ was obtained using 5000 plane waves to describe the wave functions and 550 bands. The size of the dielectric matrix was 1471 $\bfG$ vectors.
For the self-energy, wave functions were described with 4000 plane waves and 820 bands were used to evaluate $\Sigma_{\rm xc}$.

For computationally heavier QSGW calculations, 
the basis set had to be reduced introducing an error bar of 0.2 eV with respect to fully converged G$^0$W$^0$ results.
The parameters used are, for the screening: a matrix size of 1100, 340 bands and a cutoff of 1200 plane wave for the wavefunctions. For the self-energy calculation: 420 bands were used as well as a cutoff of 1200 plane waves for the wave functions.

One delicate point in the evaluation of the GW self-energy is frequency integration. Since $\Sigma_{\rm xc}$ is a product of $G$ and $W$ in real space and time, it becomes a convolution in frequency space,
\begin{equation} 
\Sigma_{\rm xc}(\rv,\rv',\omega) = \frac{i}{2\pi} \int d\omega'\; e^{i\eta\omega'}G(\rv,\rv',\omega+\omega')W(\rv,\rv',\omega').  
\label{conv}
\end{equation} 
We have performed the frequency integration using the Godby-Needs plasmon-pole model (PPM) \cite{Godby1989}  for the frequency dependence of the inverse dielectric function. The model reads
	\begin{equation}
	\epsilon^{-1}_{\gv\gv'}(\qv,\omega) = \delta_{\gv\gv'} +  \frac{A_{\gv\gv'}^2(\qv)}{\omega^2 -  (\omega^p_{\gv\gv'}(\qv) -i\eta)^2}
	\end{equation}
	 where A and $\omega^p$ are parameters that are fitted to two RPA calculations of $\epsilon^{-1}$, one for $\omega=0$ and one for a frequency on the imaginary axis, of order of the plasmon frequency. This fit is done for every $(\qv,\gv,\gv')$ element of $\epsilon^{-1}$.  In this way,
the frequency integration in Eq. \eqref{conv} is done analytically.
We have validated the PPM results with respect to those obtained with
the accurate contour deformation 
technique\cite{PhysRevB.67.155208},
where one chooses a contour in the complex plane that yields the result of the frequency integral Eq. \eqref{conv} in form of a sum over residues plus  an integration on the imaginary frequency axis.

\subsection{Optical absorption with the Bethe-Salpeter equation}
\label{subsec:bse}

The electron addition and removal quasiparticle band structure obtained from the GW calculation can be used as starting point to determine the linear response properties in the framework of MBPT, as an alternative to TDDFT. The density-density response function $\chi$ is linked to the two-particle correlation function $L$ by the relation
\begin{equation}
    \chi(\rv_1,\rv_2;t_1-t_2) = -i
     L(\rv_1,t_1,\rv_1,t_1,\rv_2,t_2,\rv_2,t_2).
\end{equation}
$L$, in turn, can be obtained from the solution of the BSE \cite{Bethe1951,Strinati1988}. In the GWA  and neglecting variations of the screening upon perturbation of the system this equation reads
\begin{multline}
L(1,2,3,4) =  L^0 (1,2,3,4) \\ +  L^0 (1,2,\bar 5,\bar 6)\left[ v(\bar 5,\bar 7)\delta(\bar 5,\bar 6)\delta(\bar 7,\bar 8) \right. \\ 
\left. - W (\bar 5,\bar 6)\delta(\bar 5,\bar 7)\delta(\bar 6,\bar 8)\right] L(\bar 7,\bar 8,3,4).
\label{bse}
\end{multline}
Here, $(1)$ is a shorthand notation for position, time, and spin $(\rv_1,t_1,\sigma_1)$, barred indices are integrated over. $L^0 (1,2,3,4)=G(1,3)G
(4,2)$ is the two-particle correlation function in absence of interaction between the two particles, and $W$ is the screened Coulomb interaction calculated within RPA. As before, we will not consider spin in the following. 

As a further approximation, usually the quasiparticle approximation \eqref{eq:qpapp} is made for $G$ in $L_0$ and 
the frequency dependence of $W$ is neglected in the kernel of the BSE. In this case, one can immediately set $t_1=t_4 $ and $t_2=t_3$ in Eq. (\ref{bse}), and the resulting equation can be reformulated as an eigenvalue problem with an effective electron-hole hamiltonian $H_{\rm exc}$, where $v_c$ and $W$ show up as effective electron-hole interactions   \cite{Hanke1979,Albrecht1998,Rohlfing2000,Onida2002}. This hamiltonian is usually expressed in a basis of pairs of orbitals. In systems with a gap at zero temperature, only pairs of an occupied and an unoccupied orbital contribute to the absorption spectrum, so the pair corresponds to a transition $|t\rangle$.  In this basis the hamiltonian reads
\beq
\bra{t} H_{\textrm{exc}}\ket{t'}= 
E_t\delta_{t,t'} + \bra{t} v_c-W \ket{t'}
\label{eqBSE}, 
\eeq
where the energy $E_t$ is the difference between an unoccupied and an occupied quasiparticle state, calculated in the GWA, and
\begin{widetext}
\begin{equation}
\bra{t} v_c \ket{t'} =  \bra{n_1\bfk_1n_2\bfk_2} v_c \ket{n_1^\prime\bfk_1^{\prime}n_2^\prime\bfk_2^\prime} = 2 \int d\mathbf{r}d\mathbf{r}^\prime
\phi^*_{n_2\mathbf{k}_2}(\mathbf{r})\phi_{n_1\mathbf{k}_1}(\mathbf{r}) v_c(\mathbf{r},\mathbf{r}^\prime)  
\phi_{n_2^\prime\mathbf{k}_2^\prime}(\mathbf{r}^\prime
)\phi^*_{n_1^\prime\mathbf{k}_1^\prime}(\mathbf{ r}^\prime),
\label{eqv}
\end{equation}
\begin{equation}
-\bra{t} W \ket{t'} = -\bra{n_1\bfk_1n_2\bfk_2} W \ket{n_1^\prime\bfk_1^{\prime}n_2^\prime\bfk_2^\prime} = -\int d\mathbf{r}d\mathbf{r}^\prime
\phi^*_{n_2\mathbf{k}_2}(\mathbf{r})\phi_{n_2^\prime\mathbf{k}_2^\prime}(\mathbf{r})W(\mathbf{r},\mathbf{r}^\prime)
\phi_{n_1\mathbf{k}_1}(\mathbf{r}^\prime)\phi^*_{n_1^\prime\mathbf{k}_1^\prime}(\mathbf{r}^\prime)
\label{eqW}
\end{equation}
\end{widetext}
are, respectively, matrix elements of the repulsive electron-hole (e-h) exchange  interaction and of the direct e-h interaction, which is usually attractive. Here, we have defined the transitions $t: (n_1 \kv_1) \to (n_2 \kv_2)$. For a given $\qv$, only wavevectors $\kv_1$ and $\kv_2$ that differ by $\qv$ contribute to $\chi(\qv)$.   Therefore, the basis is made of 
resonant transitions $(v,\kv-\qv) \to (c,\kv)$ and 
antiresonant transitions $(c,\kv) \to (v,\kv + \qv)$, and
$H_{\textrm{exc}}$ takes a block matrix form:
\begin{equation}
H_{\textrm{exc}}=
\left(
\begin{array}{cc}R &K^{R,A}\\
K^{A,R} & A 
\end{array}
\right),
\label{excmat}
\end{equation}
with the resonant matrix $R$, the anti-resonant $A$, and the coupling elements $K$.
In the optical limit $\qv\to 0$, $A=-R^*$ and $K^{A,R}=-[K^{R,A}]^*$. The diagonal blocks $A$ and $R$ are hermitian and the coupling blocks $K$ symmetric. Therefore, neglecting the coupling terms is a significant simplification; this is the  Tamm-Dancoff approximation\cite{Tamm1945,Dancoff1950} (TDA). It is usually a good approximation for absorption spectra of solids. We have verified that this is also true for AgCl, 
and all results shown in the following are obtained within the TDA. 

The solution of the eigenvalue problem $ H_{\textrm{exc}}A_\lambda=E_{\lambda}A_\lambda $ yields the elements of $L$ needed to derive $\chi$, and from this $\epsilon_M^{-1}$. 
In the TDA the result reads:
\beq
\epsilon_M^{-1}(\w) =  1 + \lim_{\qv\to 0}\frac{8\pi}{N_k\Omega_0 q^2} \sum_\lambda \frac{\Big|\sum_{t} A_\lambda^{t} \tilde{\rho}_t(\qv) \Big|^2 }{\w-E_\lambda + i\eta},
\label{spectrumBSE}
\eeq
where $\tilde{\rho}_t(\qv)$ are the same oscillator strengths of Eq.\eqref{epsnlf}.

The macroscopic dielectric function can also be calculated directly in the TDA as 
\beq
\epsilon_M(\w) =  1 - \lim_{\qv\to 0}\frac{8\pi}{N_k\Omega_0 q^2} \sum_\lambda \frac{\Big|\sum_{t} \bar A_\lambda^{t} \tilde{\rho}_t(\qv) \Big|^2 }{\w-\bar E_\lambda + i\eta},
\label{spectrumBSE2}
\eeq
where $\bar A_\lambda$ and $\bar E_\lambda$ are solutions of a modified $H_{\rm exc}$, where 
the bare Coulomb interaction $v_c$ of the electron-hole exchange 
does not have its long-range component $v_c(\gv=0)$;  note that the sets of $E_{\lambda}$ and $\bar E_{\lambda}$ contain both positive and negative energies.  
They are typically shifted with respect to the independent-particle transition energies $E_t$. If the exciton energy $\bar E_\lambda$ is smaller than the direct gap (i.e. the smallest $E_t$), then the exciton is said to be bound and the difference $E_t - \bar E_\lambda$ is its binding energy.
The coefficients $\bar A_\lambda$ mix the previously independent transitions contained in $\tilde \rho$, which can be seen from comparison with Eq. \eqref{epsnlf}.

This comparison suggests to analyze spectra in terms of the independent transitions that contribute to a given many-body transition $\lambda$. 
 The eigenvectors of the excitonic hamiltonian, $|\bar A_{\lambda}^{t}|^2$ as a function of $t$ or $E_t$, indicate how much each transition between an occupied and an empty state is mixed into the excitonic eigenstate $\lambda$. 
The electron-hole correlation in real space can be examined by investigating the
e-h wavefunction,
\beq
\Psi_\lambda(\bfr_h,\bfr_e) = \sum_{t} \bar A_\lambda^{t}\phi^*_{v\bfk}(\bfr_h)\phi_{c\bfk}(\bfr_e).
\label{excwf}
\eeq
In particular, one can fix the position $\bfr_h=\bfr_h^0$ of the hole and visualize the corresponding density distribution of the electron, $n(\bfr_e)=|\Psi_\lambda(\bfr_h^0,\bfr_e)|^2$.

Finally, the partial sum over a transition range
\begin{eqnarray}
\left|\sum_{t=1}^{t_{\rm max}} \bar A_{\lambda}^{t}\tilde \rho_t\right|^2
\label{eq:partial-sum}
\end{eqnarray}
takes the phase of the coefficients and matrix elements into account.  
As a result of positive or negative interference effects, the exciton $\lambda$ contributes more or less to the absorption spectrum. If the value of Eq. \eqref{eq:partial-sum}   for $t_{\rm max}\rightarrow\infty$ is negligibly small, the exciton is said to be dark. Otherwise, if it has a significant contribution to the absorption spectrum the exciton is called bright.

All BSE calculations have been performed using the EXC code \cite{EXCcode}.
As input,  we use the KS band structure corrected by a scissor shift taken from our GW calculation. We have verified that this reproduces  well the effect of the true GW corrections.
In the following we will refer to G$^0$W$^0$+BSE or QSGW+BSE for BSE calculations that use a scissor determined from G$^0$W$^0$ or QSGW band structure calculations, respectively.
The results presented in this paper have been obtained with the following parameters:  i) the full G$^0$W$^0$+BSE calculation has been done with 2084 shifted\cite{Benedict1998} $\kv$-points, 8 occupied and 6 unoccupied bands;  ii) both  G$^0$W$^0$+BSE and QSGW+BSE calculations with model screening (see Sec. \ref{sec:model}) have been performed using  6912 shifted $\kv$-points instead of 2084.
To obtain the  spectra $\varepsilon_M(\w)$, we have used the Haydock iterative scheme
\cite{Haydock1980,Benedict1998a}, instead of generating  eigenvalues and eigenvectors;
iii) all exciton analysis instead has been done by full diagonalization of the excitonic hamiltonian using  a four-time shifted $6\times6\times6$  Monkhorst-Pack $\kv$-point grid, 8 occupied bands and 6 unoccupied bands.

 \section{Results}
\label{sec:results}

\subsection{Kohn-Sham band structure}
\label{sec:bandstructure}

AgCl crystallizes in the fcc rocksalt structure. Calculations are carried out at the room-temperature experimental lattice constant\cite{Berry1955,Hull1999}  $a_{\rm exp}=5.55$ \AA.

\begin{figure}[!t]
  \includegraphics[width=8cm]{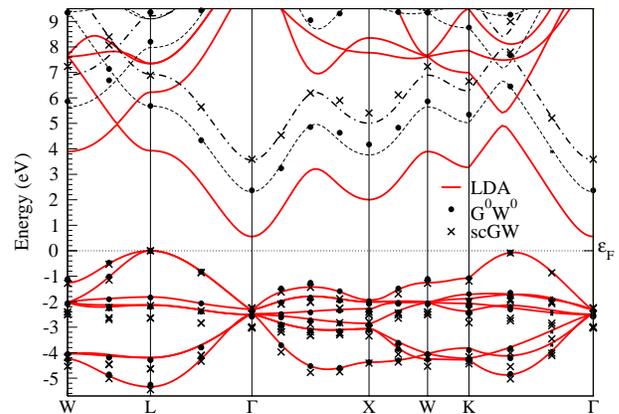}
    \caption{Calculated band structure of silver chloride. The top valence energy has been aligned to zero in all cases. 
    Red lines are the LDA calculation, the dots the  G$^0$W$^0$ results and the crosses the QSGW results. The dashed and dot-dashed lines represent the conduction bands shifted from the LDA by 1.9 eV and 3.0 eV,  respectively.    \label{fig:Band_structure}
}
\end{figure}

The Kohn-Sham band structure in the LDA is shown in Fig. \ref{fig:Band_structure}.
With the top valence at $L$ and the bottom conduction at $\Gamma$, the minimum gap is indirect and amounts to 0.56 eV.
The minimum direct gap lies close to $\Gamma$, at about 2/9 $\Gamma-K$, followed by a direct gap at 1/6 $\Gamma-X$ and the direct gap at $\Gamma$; these gaps are 2.78, 2.85 and 2.86 eV, respectively. 
These values are consistent with previous LDA calculations \cite{Ma2012,Benmessabih2007,Okoye2002,deBoer1999,Vogel1998,Gao2018}.
In GGA the indirect $L \rightarrow \Gamma$  gap is 0.3 eV larger 
\cite{Okoye2002,Zaoui2005,Benmessabih2007,Amrani2007,vanSetten2017,Zhang2019}.

Since, to the best of our knowledge, inverse photoemission spectra are not available for AgCl, the fundamental gaps have been extracted from optical measurements\cite{Ueta1986} and resonant Raman scattering experiments\cite{Nakamura1983,vonderOsten1984}.
The lowest-energy peak in the absorption spectrum gives an estimation of the minimum direct gap. The tail of the peak extending towards low energies is due to  phonon-assisted absorption processes\cite{Ueta1986,vonderOsten1984}: its edge can be used to infer the value of the indirect gap.
However, in both cases one also has to take into account the fact that the photoemission gap is larger than the absorption peak position by the exciton binding energy, whose estimation can introduce uncertainties in the band gap value. We will come back to this point in Sec. \ref{sec:excitons}.
In any case, the KS gaps severely underestimate the experimental values of 5.13 eV  for the direct optical gap \cite{Carrera1971}, and 3.25 eV for the indirect absorption edge  \cite{Brown1961,Nakamura1983}.

Earlier theoretical studies\cite{Bassani1965,Tejeda1975,Kunz1982,vonderOsten1984,Ueta1986} have found that Ag$^+$ $4d$ and Cl$^-$ $3p$ ionic states have similar energy in the crystal, leading to strong hybridization in the valence band. While  
their mixing is zero at $\Gamma$, it is strong elsewhere, notably at $L$. This $\bfk$-dependent hybridization and the strong $p-d$ repulsion cause the upward curvature of the top-valence bands at $\Gamma$ (i.e. a negative hole effective mass) and make AgCl an indirect semiconductor.
On the contrary, in the alkali halides, which share the same rocksalt crystal structure, the ionic energy levels are well separated, leading to a much larger ionic character of the compounds and  a direct band gap. 

\begin{figure}[!t]
    \centering
    \includegraphics[angle=270,width=\columnwidth]{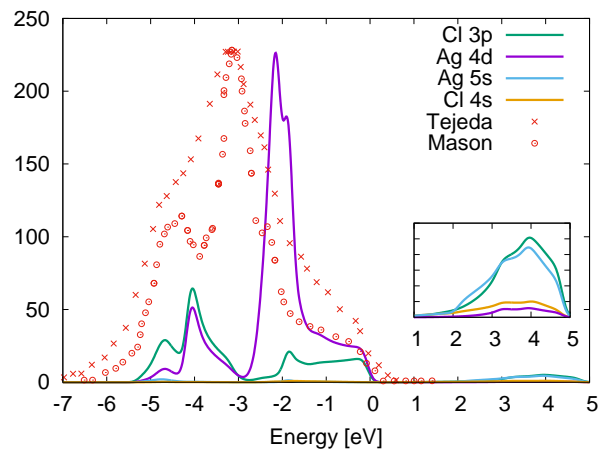}
    \caption{Projected density of states of silver chloride calculated in KS-LDA compared to photoemission spectra from  Mason\cite{Mason1975} and Tejeda {\it et al.}\cite{Tejeda1975} at a photon energy $h\nu = 1486.6$ eV. 
    In each curve the top valence has been aligned to zero and the intensity scaled to the maximum of the most prominent peak. In the inset: zoom on the unoccupied PDOS.}
    \label{fig:dos}
\end{figure}

Our calculations, as shown by the band structure in Fig. \ref{fig:Band_structure} and the projected density of states (PDOS) in Fig. \ref{fig:dos},  confirm this picture. 
While Cl $3s$ states are located at $\sim$ -15 eV (not shown), the valence band region comprises 8 bands.  
They are very close to each other at the $\Gamma$ point, where from the bottom to the top we count 3 degenerate Ag $t_{2g}$ states\footnote{The crystal field at Ag site has a cubic point symmetry ($O_h$).}, 2 degenerate Ag $e_{g}$ states, and 3 degenerate Cl $3p$ states. 
The hybridisation between  Ag $4d$   and Cl $3p$ increases moving away from the $\Gamma$ point towards the top and the bottom of the valence bands at the $L$ point, where  Ag $4d$ and  Cl $3p$ are strongly mixed. Their interaction gives rise to dispersive bands.
Instead, three Ag $4d$ bands, which are not dispersive between $\Gamma$ and $L$, remain at the center of the valence manifold, giving rise to a pronounced sharp peak in the PDOS. 
Finally, the lowest conduction band 
has a delocalised Ag $5s$ and Cl $4s$ character at the $\Gamma$ point \cite{deBoer1999} and mixed Ag $5s$ - Cl  $3p$ elsewhere. 


In Fig. \ref{fig:dos} the calculated PDOS is compared to experimental photoemission spectra measured at 1486.6 eV photon energy by Mason\cite{Mason1975} and Tejeda {\it et al.}\cite{Tejeda1975}. The two measured spectra, taken at the same photon energy, differ in their shapes, illustrating the experimental difficulty of the characterisation of the electronic properties of AgCl. Still we can analyse their common features. 
The measured spectra are characterised by a band width of about 6 eV and a main peak centered at about -3.1 eV.
 The calculation correctly describes the presence of shoulders about 4 eV below the main peak and about 2 eV above it.
 On the basis of atomic photoionization cross sections\cite{Yeh1985}, we find that photoemission spectra  at $h\nu =1486.6$ eV mostly probe the Ag $4d$ electrons. 
We can therefore directly compare the experimental spectra to the calculated  Ag $4d$ PDOS. 
We thus assign the main peak to the nondispersive Ag $4d$ bands,  which result too shallow by about 1 eV in KS-LDA. 
This underestimation of the binding energy of occupied localised $d$ levels is a common tendency of KS-LDA  
that can be improved by the GWA (see e.g. \cite{Bechstedt2014,Wei1988,Kotani2007,Christensen2010,Svane2011,Grueneis2014}), as we will discuss more in detail in the next section.

\subsection{Band structure in the GW approximation}

The band structure of silver chloride evaluated in G$^0$W$^0$ has been added to the KS-LDA band structure in  Fig.~\ref{fig:Band_structure}.
The top of the valence band is aligned to zero.
The G$^0$W$^0$ band structure is similar to the KS-LDA one, besides an almost rigid shift of 1.8-1.9 eV of the conduction bands. 
For illustration, the dashed curve shows the  lowest conduction bands in KS-LDA, shifted upwards by 1.9 eV.

This size of the band-gap opening is in agreement with the value of 1.75 eV recently obtained by Zhang and Jiang \cite{Zhang2019} in a G$^0$W$^0$ full-potential linearized
augmented plane wave calculation starting from a KS-GGA band structure. 
Instead, van Setten {\it et al.}\cite{vanSetten2017} reported a much smaller G$^0$W$^0$ correction  to the KS-GGA gap: 1.25 eV. 
The reason of this large discrepancy should not be ascribed to inaccuracies in the pseudopotentials \cite{Zhang2019}, 
but rather to an underconvergence problem: their automatized algorithm employed only 155 bands (compared to 820 in the present work).
On the contrary, Gao {\it et al.}\cite{Gao2018} found a much larger G$^0$W$^0$ correction  starting from KS-LDA, i.e. 2.38 eV, 
and had to include up to 2500 empty bands in the  G$^0$W$^0$ calculation.
A similar situation was previously encountered in other materials like ZnO\cite{Shih2010,Friedrich2011,Stankovski2011} and TiO$_2$\cite{Kang2010}, where semicore electrons have to be explicitly included in the GW calculation. The origin of the problem  in those calculations was identified\cite{Kang2010,Stankovski2011} with the use of the $f$-sum rule in the Hybertsen-Louie  \cite{Hybertsen1986} PPM, which was adopted also by Ref. \cite{Gao2018} for AgCl.
Indeed, also in the present case our results obtained with the Godby-Needs PPM agree (within 0.2 eV at most) with the accurate contour-deformation (CD) calculation that avoids any PPM (see Tab. \ref{tab:Comp}). 
As a final validation, we have also employed the effective energy technique  \cite{Berger2010} (EET) that accounts approximately for all empty states and allows reaching convergence much more easily that the traditional sum-over-states scheme. Using the EET (here used within the PPM) the values for the band gaps are once again in agreement within 0.1 eV (see Tab. \ref{tab:Comp}).

\begin{table}[!t]
    \caption{Direct and indirect photoemission gaps from different approximations compared to experimental (Exp.) absorption onsets from optical measurements (Refs. \onlinecite{Brown1961,Carrera1971}) which provide a lower bound due to excitonic effects (see Sec. \ref{sec:excitons}).}
        \label{tab:Comp}
    \centering
    \begin{ruledtabular}
    \begin{tabular}{ccc}
          & Indirect & Direct \\
          \colrule
         LDA & 0.56 &  2.78  \\
         G$^0$W$^0$ (PPM) & 2.4  &  4.8 \\
         G$^0$W$^0$ (CD)  &   2.4 & 4.6  \\
         G$^0$W$^0$ (EET) & 2.4 & 4.7\\
         QSGW     & 3.7  &  5.9\\
         evQSGW     & 3.2  & 5.7 \\
         Absorption onset (Exp.) & 3.25 & 5.13 \\
    \end{tabular}
    \end{ruledtabular}
\end{table}

The G$^0$W$^0$ indirect band gap is now 2.4 eV and the 
 direct gap at $\Gamma$ is 4.6-4.8 eV. Both are still smaller than the experimental optical gaps (see Tab. \ref{tab:Comp}).
However, in situations with large $pd$ hybridisation as for AgCl the LDA starting point may not be reliable \cite{Schilfgaarde2006}.
On the other hand, also the large corrections obtained within the G$^0$W$^0$ scheme question the first-order perturbative approach itself.

In order to overcome the problem of the KS-LDA starting point and assess the  G$^0$W$^0$ perturbative scheme, 
we have performed QSGW calculations. The new band structure is shown in Fig.~\ref{fig:Band_structure}; again, top-valence bands are aligned. At first sight, there is no drastic change in the dispersion of the valence and conduction bands. However, a closer look shows that, whereas the G$^0$W$^0$ valence bands were essentially on top of the KS-LDA ones, QSGW results slightly increase -- by 0.1 eV -- the valence bandwidth and push the narrow Ag $4d$ bands down by 0.5 eV, leading to a better agreement with photoemission results\cite{Mason1975,Tejeda1975}.  
The most obvious change is the almost rigid shift of the conduction bands with respect to KS-LDA, which has passed from 1.9 eV in G$^0$W$^0$ to 3.0 eV, as indicated by the dot-dashed line in Fig.~\ref{fig:Band_structure}. 

With this shift, the indirect gap becomes 3.7 eV while 
the direct band gap is 5.9 eV. The fact that band gaps seem to be overestimated (see Tab. \ref{tab:Comp}) may have two reasons. First, the self-consistent RPA screening in QSGW is too weak, which brings results too close to Hartree-Fock \cite{Schilfgaarde2006,Svane2011,Bruneval2014,the-book}.  
Second, the experimental optical gaps are affected by excitonic effects (see Sec.~\ref{sec:excitons}).

We have also performed a QSGW calculation where only the QP eigenvalues are calculated self-consistently, while the QP wavefunctions are constrained to remain the KS-LDA orbitals. This further calculation is named 'evQSGW' in Tab. \ref{tab:Comp}. It gives band gaps that are intermediate between the  G$^0$W$^0$ and the full QSGW results, illustrating the impact of the change of the wavefunctions on the band structure.
Finally, we have tested the effect of the update of the screened Coulomb interaction $W$: a QSGW calculation, in which we keep the $W$ fixed at the level of the RPA-LDA, the resulting gap (direct 4.8 eV and indirect 2.7 eV) is closer to the G$^0$W$^0$ value than to the QSGW one. 
In the QSGW calculation in AgCl the modification of the screened interaction $W$ is hence the most critical effect.

\subsection{Absorption spectra}

Since the optical properties of AgCl are of utmost importance for its applications, their calculation and analysis represents the focus of the present work.

\subsubsection{Absorption spectrum in time-dependent density functional theory}
\label{sec:abs}

\begin{figure}[ht]
    \centering
    \includegraphics[width=\columnwidth]{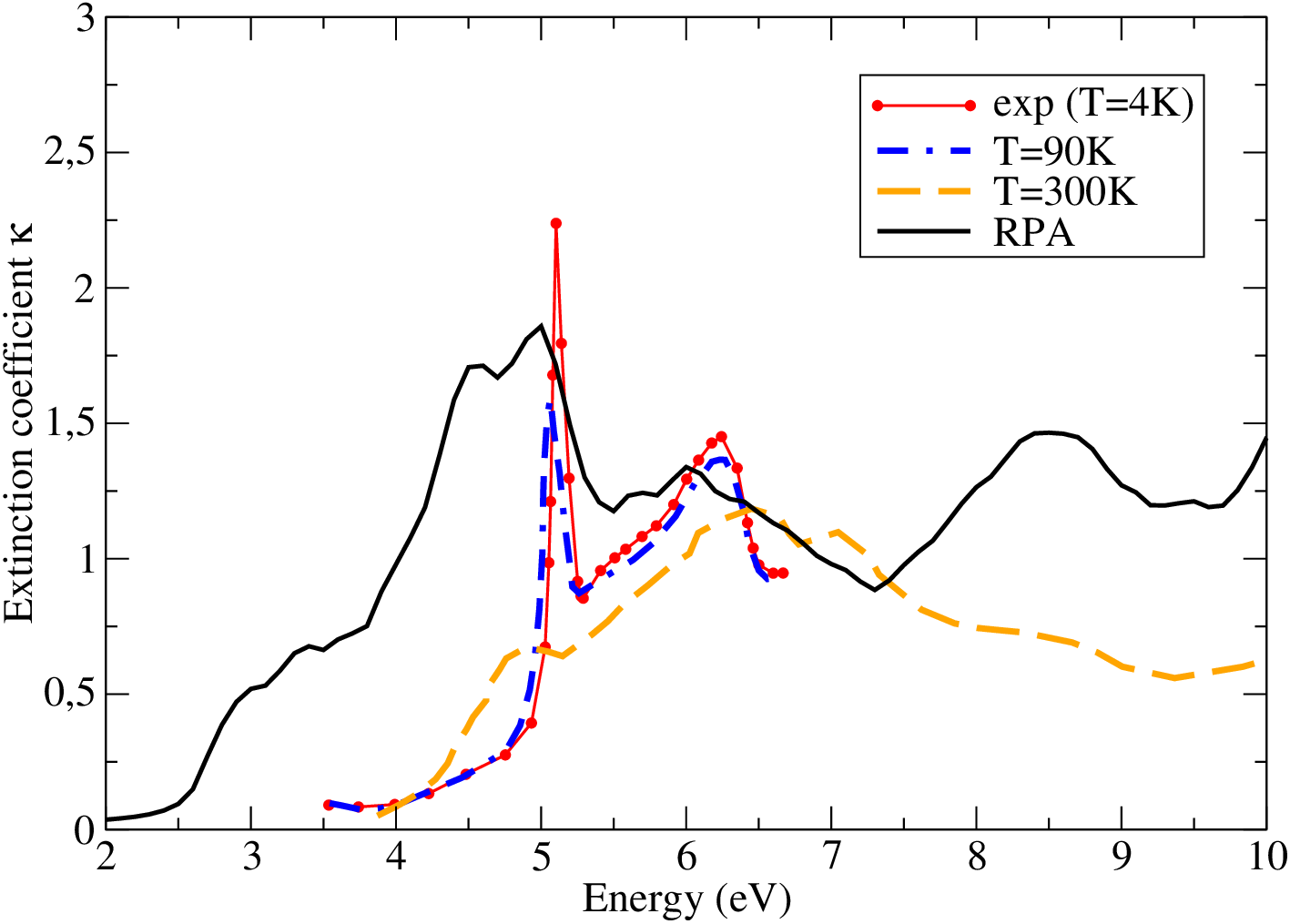}
    \caption{Extinction coefficient as a function of energy. 
    The RPA calculation based on the KS-LDA band structure (black solid curve) is compared to experiment data from Ref. \onlinecite{Carrera1971} measured at 4 K (red curve), at 90 K (blue curve) and room temperature (orange curve).}
    \label{fig:Abso_rpa}
\end{figure}

\begin{figure}[t]
    \centering
    \includegraphics[width=\columnwidth]{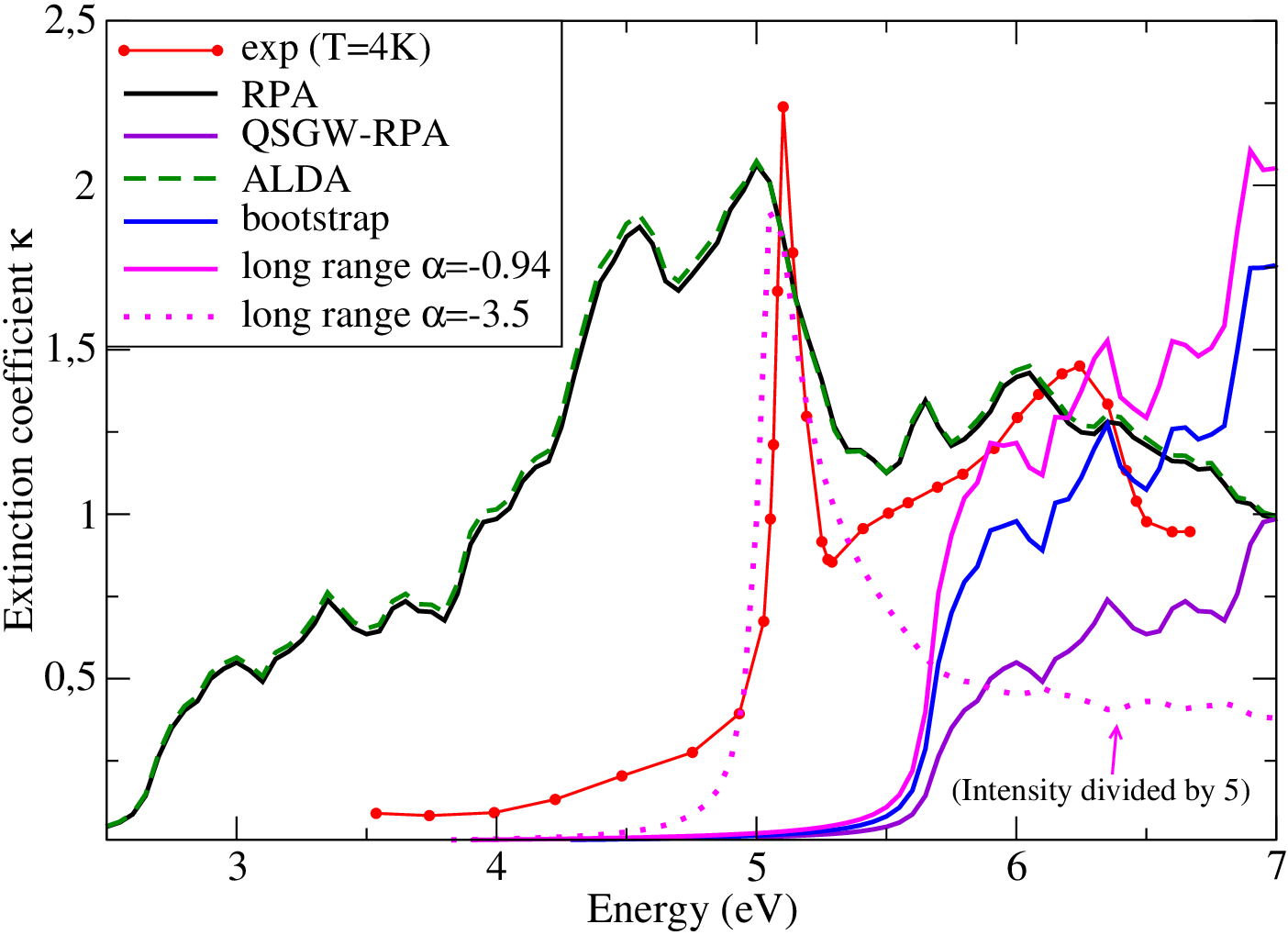}
        \caption{Comparison between the extinction coefficient measured at  4 K with several TDDFT approximations. Since the energy range is limited to 3-7 eV, these spectra have been calculated with only 4 conduction bands.}
    \label{fig:tddft_agcl}
\end{figure}

 Fig. \ref{fig:Abso_rpa} shows the extinction coefficient $\kappa(\w)=\text{Im}\sqrt{\varepsilon_M(\w)}$. 
 We compare the RPA result (black curve) 
 with three experiments at different temperatures: 4 K (red curve), 90 K (blue curve), 300 K (orange curve). The spectra have been measured up to 6.7 eV in Ref. \onlinecite{Carrera1971}. The wider range at room temperature has been obtained by combining results from different sources. 
 The shape of the measured spectra is strongly affected by temperature: the very sharp peak at the onset of the spectrum around 5.1 eV is clearly visible only at low temperatures, while the room-temperature spectrum is much broader.
 Since our calculations do not include the effect of temperature, comparison to the low-temperature experiment should be more meaningful.
Still, keeping this fact in mind, also the room temperature experiment gives important indications.
Overall, the RPA and experimental spectrum at 300 K are similar. However, the absorption onset is underestimated in the RPA, by more than 1 eV. Moreover, 
the RPA entirely misses the sharp feature at the onset of the low-temperature experimental result.   
The underestimate of the onset is a common problem in KS-RPA spectra\cite{Onida2002}. Since $f_{\rm xc}=0$ and only short-range components of $v_c$ contribute to optical spectra, the onset is determined by the interband transitions in $\chi_0$, as can be seen from Eq.~\eqref{dysonchi} and Eq. \eqref{chi0}. It suffers therefore both from the use of the LDA, and from the fact that even the exact  KS band gap would be smaller than the measured band gap. The results do not change when $f_{\rm xc}$ is taken into account within the ALDA: in Fig. \ref{fig:tddft_agcl} the ALDA (dashed green curve)  is hardly distinguishable from the  RPA (black curve). The ALDA can neither lead to a significant opening of the optical gap, nor to a significant change in the spectral shape. 

It is well established \cite{Ghosez1997,Reining2002} that the ALDA suffers from the absence of a long-range contribution that in nonzero-band-gap materials would diverge as $1/q^2$ for the large wavelength of the light, $q\to 0$. The exact xc kernel should contain such a contribution. Several suggestions exist how to include a long-range component in $f_{\rm xc}$\cite{Reining2002,Botti2004,Trevisanutto2013, Hellgren2013}. 
All kernels of this family start from a $\chi_0$ built with a quasiparticle band structure from GW or similar approaches, instead from a KS one.
A simple static and scalar  $f_{\rm xc}$  with a long-range contribution 
can then simulate the effects of the electron-hole interaction by shifting spectral weight to lower energies.

In order to simulate the missing excitonic effects, we have examined the long-range kernel\cite{Reining2002} of the type $\alpha/q^2$, where the band-gap opening  at the QSGW level is accounted for by a scissor correction of 3.0 eV.
The result is given by the pink curve in Fig. \ref{fig:tddft_agcl}. The value of $\alpha=-0.94$ has been obtained from Eq. (4) of Ref. \onlinecite{Botti2004}  using an experimental dielectric constant $\epsilon_{\infty}=~4$. 
In contrast to the ALDA, the long-range kernel does shift the spectral weight to lower energies with respect to the QSGW-RPA result (violet curve), where $f_{\rm xc}=0$. However, the spectral onset remains the same, overestimating the experimental result, and the sharp peak is still missed. 
A larger value of $|\alpha|$ in the long-range kernel  $\alpha/q^2$
would enhance the excitonic effects. However, in order to reproduce the sharp experimental peak, we should increase the strength $|\alpha|$ to very large values, which would completely destroy the spectrum. For example, the dotted curve in Fig.  \ref{fig:tddft_agcl} is obtained with $\alpha=-3.5$. Note that its overall intensity is divided by a factor 5 in the plot, while all spectral features at higher energies have collapsed. This failure of static long-range kernels is confirmed also by similar approaches, such as the recent bootstrap kernel \cite{Sharma2011,Rigamonti2015} (see the blue curve in Fig. \ref{fig:tddft_agcl}, obtained with the implementation of Eq.(5) of Ref.\onlinecite{Rigamonti2015}).
Therefore, we can conclude that TDDFT with the presently available simple approximations does not give a good description of the optical properties of AgCl, and in particular, of the strong excitonic effects that should explain the remaining discrepancy between theory and experiment. For this reason, we have to move on to a full description in the framework of MBPT, by solving the BSE. 

\subsubsection{Optical spectra from the Bethe-Salpeter equation: insight and difficulties}
\label{sec:firstBSE}

\begin{figure}[t]
	\centering
	\includegraphics[width = \columnwidth]{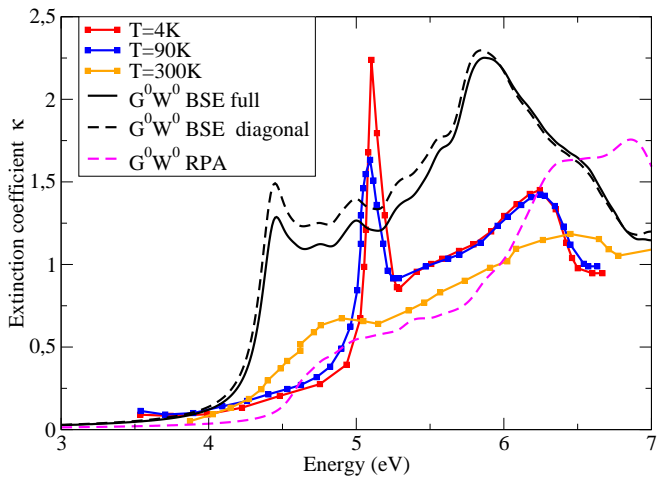}
	\caption{
		Comparison between the experimental extinction coefficient from Ref. \onlinecite{Carrera1971} at different temperatures  
		and MBPT calculations (with 2084 $\bfk$ points): G$_0$W$_0$-RPA obtained with 1.9 eV scissor correction (pink dashed line)   G$_0$W$_0$+BSE obtained with the full screening matrix $W_{\bfG\bfG'}(\bfq)$ (black solid line) or only its diagonal contribution (black dashed line).		}
	\label{exp_vs_bse}
\end{figure}

A state-of-the art BSE calculation starts from the G$_0$W$_0$ band structure (here simulated by a scissors shift of 1.9 eV) and employs the LDA-RPA screened Coulomb interaction $W$ to account for the electron-hole attraction.
This kind of calculations is 
computationally expensive. However, even a calculation with  reduced parameters can give an idea of the importance of excitonic effects. 
We therefore start by looking at the result of a G$_0$W$_0$+BSE calculation in Fig. \ref{exp_vs_bse} (obtained with 2084 $\bfk$ points). 

The G$_0$W$_0$-RPA  onset of the spectrum (pink dashed line in  Fig. \ref{exp_vs_bse}) underestimates the experimental threshold and does not show a pronounced peak at low energy: it merely shifts  the LDA-RPA spectrum to higher energy. The electron-hole interaction in the BSE shifts oscillator strength to lower energies, and a peak forms (black solid line in  Fig. \ref{exp_vs_bse}).
However, the G$_0$W$_0$+BSE spectrum   is now at an even lower energy  and the excitonic peak is much too weak with respect to experiment.

BSE calculations in solids are often done by neglecting the off-diagonal elements of the screening matrix\cite{Albrecht1998,Onida2002}  $W_{\bfG\bfG'}(\bfq)$ that represents the direct electron-hole interaction. This is justified when the electron-hole pair is delocalized enough to justify a space-averaged screening. The first excitonic peak in AgCl is influenced by this approximation: taking into account the full spatial details of screening  (solid black line in  Fig. \ref{exp_vs_bse}) reduces the peak intensity by about 10\% with respect to the approximation of diagonal screening (dashed black line in  Fig. \ref{exp_vs_bse}). 

This G$_0$W$_0$+BSE calculation remains qualitative for several reasons. First, as pointed out above, the G$_0$W$_0$ band structure from an LDA starting point is not reliable for AgCl. Second, the spectrum is also strongly dependent on the Brillouin zone sampling, and a $\bfk$-point convergence test performed with a reduced number of conduction bands shows that a set of 6912 $\bfk$ points is needed instead of 2084 employed here.

While the first issue could be solved by using the QSGW band structure as a starting point for the BSE calculation, 
the main computational problem would still remain the setting up of the full screening matrix $W_{\bfG\bfG'}(\bfq)$ that should be calculated self-consistently for too many $\bfq$ points.

In order to overcome this problem, we will complement the first principles calculations with a model screening, where the parameters of the model are fitted to the \textit{ab initio} results. As we will see in the next section, a careful analysis allows us to turn this simple approach  into a powerful way to obtain reliable results. 

\section{Model screening of the electron-hole interaction}\label{sec:model}

The screened Coulomb interaction is evaluated in Fourier space as 
$$
W_{\gv\gv'}(\qvt)=\epsilon^{-1}_{\gv\gv'}(\qvt,\w=0){v_c}_{\gv'}(\qvt),
$$  
where $\qvt=\kv-\kv'$ must correspond to the difference of two $\kv$-points on the grid used in the calculations. 
\footnote{Since we are in the optical limit ($\qv\to 0$), this is equivalent, in Eq.(\ref{eqW}), to consider $\kv_1-\kv_1'=\kv_2-\kv_2'=\qvt $.} 
Therefore, with increasing $\kv$-point grid size an increasing number of screening matrices has to be calculated. Even though the screened electron-hole interaction is calculated within the RPA  and at $\w=0$, this quickly constitutes a formidable task, especially within the QSGW scheme.   

\begin{figure*}[!thb]
    \includegraphics[width=\columnwidth]{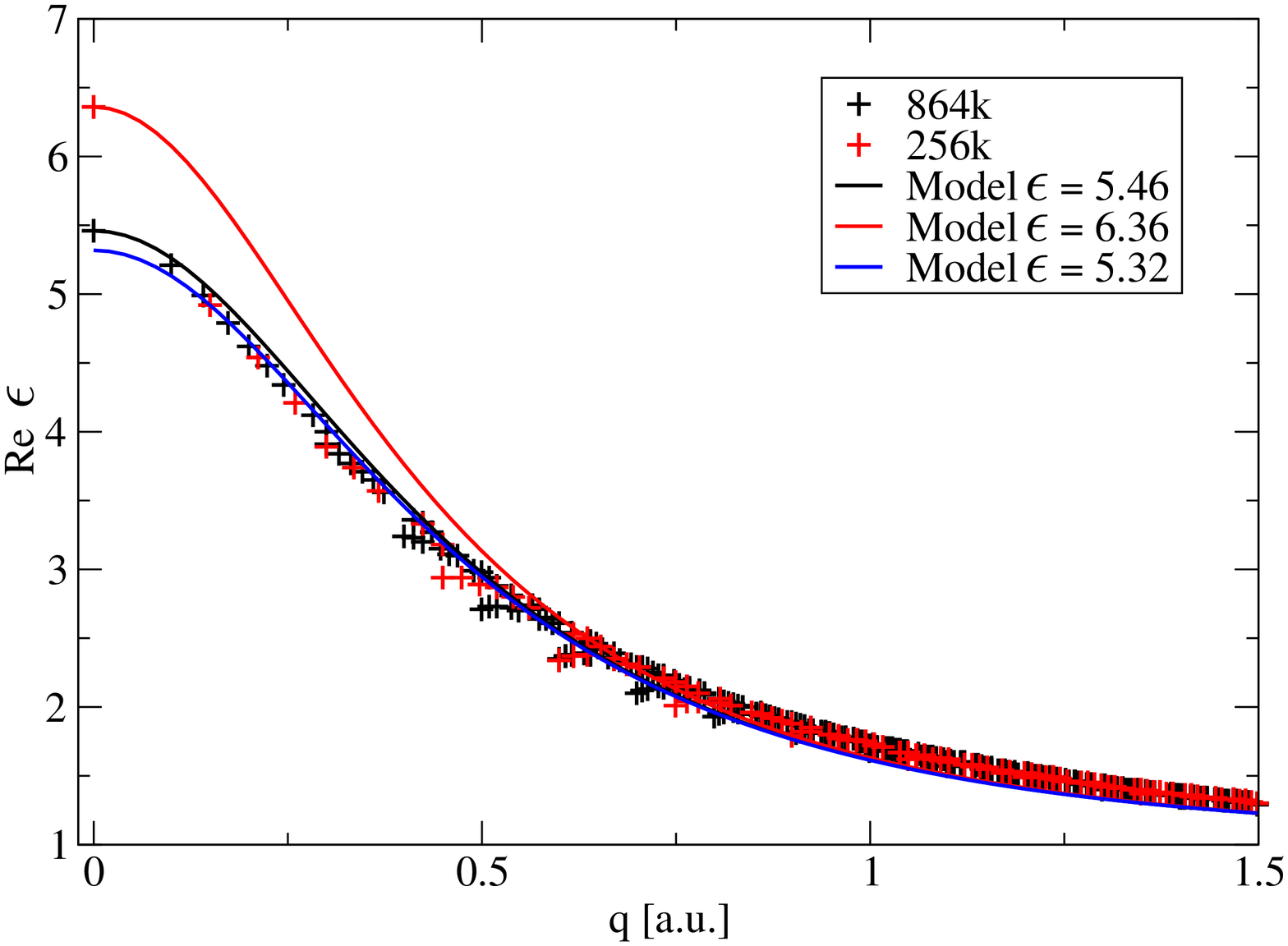} 
    \includegraphics[width=\columnwidth]{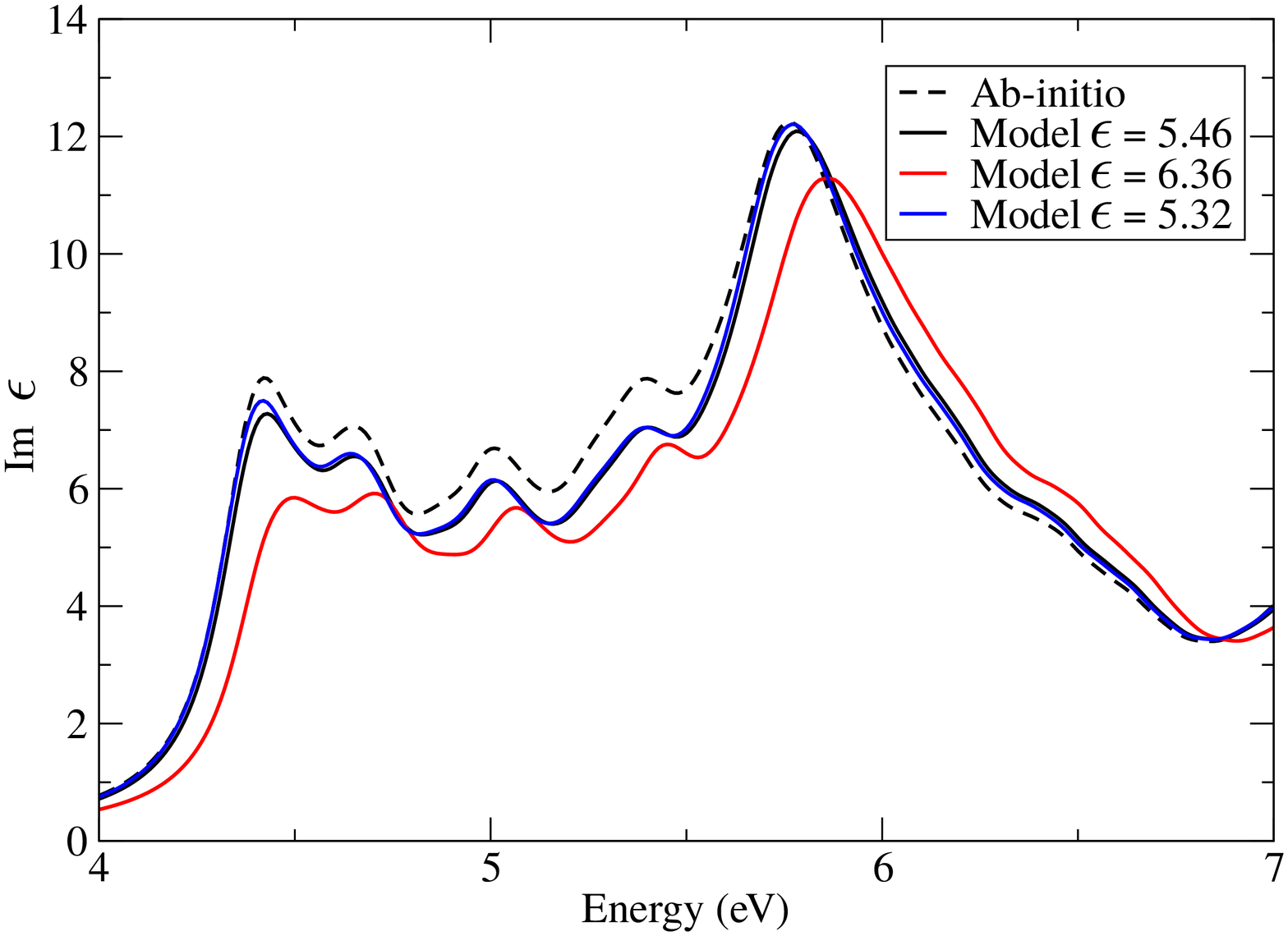}
    \caption{Effects of the model electron-hole screening on BSE results. Left panel: \textit{Ab initio} LDA-RPA (crosses) and model (solid lines) dielectric constants $\epsilon(\bfq,\w=0)$ as a function of $\qv$. The \textit{ab initio} results are obtained using 4 shifted $4\times4\times4$ (red) or  $6\times6\times6$  (black)  Monkhorst-Pack grids of $\kv$ points, corresponding to 256 or 864  $\kv$ points in the full Brillouin zone, respectively.
    The black cross at $q\to 0$  also contains the correction for the non-local pseudopotential, while the red cross does not. Model results are fitted to the \textit{ab initio} results at $q=0$ (respective color code) or at $q=0.15$ a.u. (blue), where results on the smaller grid are already converged. 
    Right panel: BSE spectra obtained with 864 shifted $\kv$-points 8 valence and 6 conduction bands. The screening of the electron-hole interaction is taken from the results shown in the left panel: either the diagonal of the \textit{ab initio} screened Coulomb interaction (dashed black), or the model screening fitted to the
    864 $\kv$-points result at $q=0$ (solid black), or to the 256 $\kv$-points result at $q=0$ (red) or at $q=0.15$ a.u. (blue). }
    \label{fig:model_vs_abinit}
\end{figure*}

In order to overcome this bottleneck, sometimes the \textit{ab initio} screening is replaced by a model \cite{Lundqvist1967,Johnson1974,Levine1982,Bechstedt1994,Shirley2006,Rohlfing2010,tal2020accurate,vinson2020revisiting,Keldysh1979,Cudazzo2011,Latini2015,Trolle2017}.
In particular, for bulk semiconductors a successful model  was proposed in Ref. \cite{Cappellini1993}.
It represents the static dielectric function as
\begin{equation}
    \epsilon(q)= 1 + \frac{1}{\frac{1}{\epsilon(q=0)-1 } + \alpha \left( \frac{q}{q_{TF}}\right)^2 + \frac{q^4}{4\omega_p}},
    \label{eq:model-eps}
 \end{equation}
where $q_{TF}= 2(3\bar{n}/\pi)^{1/6}$ and $\omega_p = \sqrt{4\pi \bar{n}}$, with $\bar{n}$ the average density. $\alpha$ is a parameter set to 1.563, following Ref. \onlinecite{Cappellini1993}. This model dielectric function only gives the diagonal in reciprocal space, but, as we will see, it is sufficient for the present purpose.

Although the model is very simple, its use requires care. First, the ``average density'' should not be the average density of all electrons, but only of those valence electrons that participate to the screening.  This difference is
well defined in a simple semiconductor such as silicon, but less obvious in materials like AgCl with electrons of different character in the valence bands. Since the model screening depends strongly on the density, comparison of the model results with various choices for the valence density to an, even not fully converged, \textit{ab initio} calculation in a few $\bfq$ points is sufficient to see that
the screening is determined by the electrons in the 8 upper valence states, i.e. the Ag $4d$ and Cl $3p$ electrons. To include some of the remaining, more tightly bound, electrons in the average density would clearly lead to overscreening.

Second, the macroscopic dielectric constant enters the model as an important parameter. Not always a reliable experimental value is known, and in a fully first-principles framework, it should be calculated.
This is a much more delicate point, as we will illustrate in the following. The left panel of Fig. \ref{fig:model_vs_abinit}  shows the KS-RPA dielectric function $\epsilon(\bfq,\w=0)$, calculated with a 4 times shifted $4\times4\times4$ grid of $\kv$ points, corresponding to 256 $\kv$ points in the full  Brillouin zone (red crosses). This calculation yields a macroscopic dielectric constant $\epsilon(q=0)=6.36$. The model curve (red line) that is obtained by using this value for $\epsilon_0$ reproduces  the \textit{ab initio} results very well at large $q$, and it is perfect for $q\to 0$ by definition, but for small to moderate $q$ the discrepancy is significant. As a consequence, when the full first principles screening of the electron-hole interaction in the BSE is replaced with this model screening, the comparison is not satisfactory, as can be seen in the right panel.

\begin{figure}
    \begin{center}
    \includegraphics[width=\columnwidth]{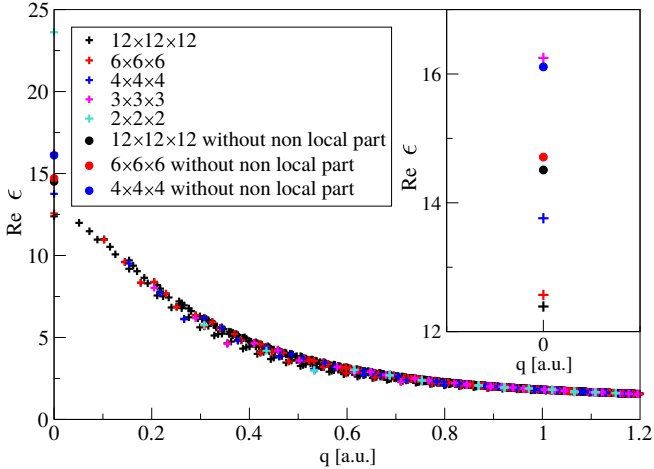}
    \caption{Bulk silicon: Static dielectric function as function of wavevector, for different $\bfk$-point grid sizes. Plus (circle) symbols are results from calculations including (excluding) the commutator with the non-local pseudopotential. In the insert: zoom on $\bfq=0$. }
    \label{fig:my_label}
    \end{center}
    \end{figure}

 Interestingly, the problem is not the model, but the \textit{ab initio} calculations: the left panel also shows the \textit{ab initio} results obtained with a denser Brillouin zone sampling (black crosses). The comparison highlights the fact that the value of the dielectric constant at $q=0$
 calculated on the coarse grid (red crosses) is not converged, contrary to the values obtained for non-vanishing $q$. Indeed, the $q=0$ calculation should be more difficult to converge  with respect to the Brillouin zone sampling: for $q\neq 0$ transition energies $\Delta \varepsilon$ enter the dielectric function in a denominator, but for  $q\to 0$
this denominator is determined by $(\Delta \varepsilon)^3$. A second difficulty is that, unless a double $\kv$-grid is used, the $q\to0$ limit requires the calculation of a correction to the dipole matrix element whenever the hamiltonian is non-local\cite{Baroni1986}. This correction, which takes the form of a commutator of the potentials with the space coordinate ${\bf r}$, gives a sometimes sizeable contribution in the case of non-local pseudopotentials, and/or when the band structure stems from a non-local self-energy\cite{Koerbel2015,Gatti2015}. Both difficulties are  
very general and even occur in simple semiconductors such as silicon.  Fig. \ref{fig:my_label} gives an illustration, with $\epsilon(\bfq,\w=0)$ of bulk silicon shown for various grids of $\bfk$ points, and including or excluding the commutator with the non-local pseudopotential. 
This is a Kohn-Sham calculation, so the pseudopotential is the only non local component.

Since by increasing the number of $\bfq$ points the contribution from $q=0$ to a sum over the Brillouin zone vanishes, the poor quality of $\epsilon(q= 0)$ is often acceptable when the \textit{ab initio} dielectric function is used in an integral, e.g., in GW or BSE calculations. Instead, when the macroscopic dielectric constant is used to determine the model dielectric function, a bad estimate of $\epsilon(q=0)$ deteriorates the model screening over a large range of wavevectors, as can be seen from the left panel of Fig. \ref{fig:model_vs_abinit} for AgCl, by comparing the fits obtained on the converged and unconverged \textit{ab initio} calculations.

One might, of course, improve the calculation of the dielectric constant, but this would reduce significantly the computational gain of using the model, especially when self-consistent QP results are used (see below). Here, we propose an alternative route, namely, we fit the model to the calculated dielectric constant at a non-vanishing momentum transfer, $q^0\neq 0$. In this way, no commutator has to be evaluated, and we can make use of the fact that $\epsilon(q^0)$ converges more quickly than $\epsilon(q=0)$. 
The choice of the $q^0$ where the model parameters should be determined is constrained: for larger $q$, and taking into account crystal local field effects, the local anisotropy of the crystal induces a scattering around the function $\epsilon(q)$ and therefore some arbitrariness. One therefore has to choose a value   $q^0$  that is small enough to yield a well defined  $\epsilon(q^0)$ and large enough to converge fast with the $\kv$-point grid. In any case, for any $q^0\neq 0$ the problem of the commutator is avoided. Eq. \eqref{eq:model-eps} now turns into
\begin{equation}
    \epsilon(q)= 1 + \frac{1}{\frac{1}{\epsilon(q^0)-1 } + \alpha  \frac{(q^2-(q^0)^2)}{q_{TF}^2}+ \frac{(q^4-(q^0)^4)}{4\omega_p}}.
    \label{eq:model-eps-2}
 \end{equation}
This approach, as we will show subsequently, is very powerful to converge optical spectra calculated from the BSE. As a byproduct, once the model parameters are determined this allows one also to extrapolate the dielectric constant at $q\to 0$. This is demonstrated in the left panel of Fig. \ref{fig:model_vs_abinit}: the model dielectric function obtained from the fit to the \textit{rough} first-principles calculation at $q^0=0.15$ a.u. (blue curve) compares very well to the one fitted to the best first-principles calculation at $q=0 $ (black curve); from the fit at $q^0=0.15$ a.u. to the \textit{unconverged} calculation, one can read an RPA dielectric constant $\epsilon(q=0)=5.32$, which well compares with the  converged \textit{ab initio} result $\epsilon(q=0)=5.46$, and which is much better than the result  $\epsilon(q= 0)=6.36$ of the unconverged \textit{ab initio} calculation itself.

\begin{figure}
    \centering
    \includegraphics[width=\columnwidth]{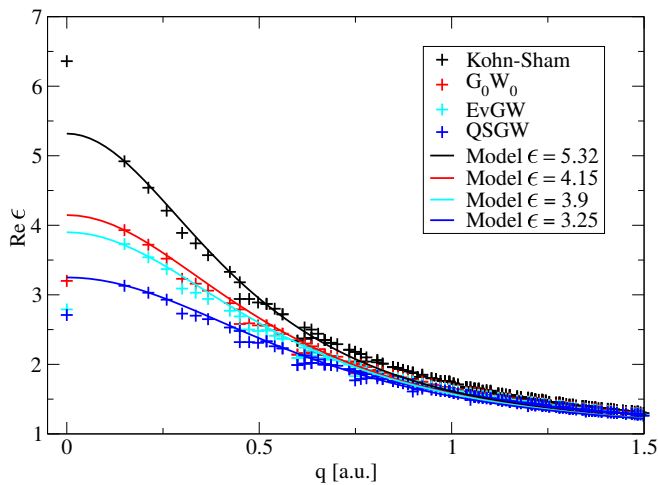}
    \caption{Silver chloride: static dielectric constant as function of momentum transfer calculated using different levels of theory: Kohn-Sham LDA,  G$^0$W$^0$, evQSGW and QSGW. All calculations are done using a 4 times shifted $4\times4\times4$ Monkhorst-Pack $\bfk$-point grid and 340 bands. The commutator with the non-local potentials (pseudopotential and self-energy, when applicable) is neglected. The fit of the model results at $\qv^0$=(-0.125,-0.125,0) in units of the  reciprocal lattice ($q=$0.15 a.u.) yields the dielectric constants.
     }
    \label{fig:welement}
\end{figure}

Note that the converged RPA dielectric constant of 5.46 is larger than the experimental value, which is found to lie between 3.7 \cite{Wakamura1996} and 3.97 \cite{PhysRevB.6.4667}. Other calculations based on KS-LDA\cite{Gao2018} also find too large dielectric constants, similar to ours.
 This may be traced back to the Kohn-Sham band gap, whose influence on the dielectric constant is in the RPA not compensated by $f_{\rm xc}$. However, for consistency  we use the RPA value in order to simulate the RPA screening. Since overscreening leads to underestimation of the excitonic effects, we have  calculated the dielectric function also using G$^0$W$^0$, evQSGW and QSGW  results as input for the RPA. Results are shown in
Fig. \ref{fig:welement}. 
As expected, the strongest screening is obtained in the KS-RPA, the same result as presented in Fig. \ref{fig:model_vs_abinit}. Self-energy corrections to the eigenvalues calculated in G$_0$W$_0$ open the gap and lower the screening, with an effect that is particularly visible at smaller wavevectors.   
Self-consistency on the eigenvalues further opens the gap and lowers the screening, again with a stronger effect at smaller wavevectors. Self-consistency in the wavefunctions in QSGW enhances the trend, with an effect that is significant at small $q$, but almost negligible above $q\approx$ 1 a.u. 
These \textit{ab initio} results have been obtained using a 4 times shifted $4\times4\times4$  grid, neglecting the commutator with the non-local pseudopotential and, in the case of the GW calculations, neglecting the commutator with the non-local self-energy. 
As a result, while the differences between the various approximations are significant but reasonable and smooth for $q\neq 0$, the same does not hold for $q= 0$. In particular, the dielectric constant is clearly too small when GW ingredients are used. The error due to the neglect of the commutator with the non-local self-energy is sizeable, and of opposite sign with respect to the pseudopotential contribution. We therefore determine the screening at $q^0=0.15$ a.u., and then use Eq. \ref{eq:model-eps-2} to obtain the full $\epsilon(q)$. This allows us to extrapolate the dielectric constants at vanishing wavevector. 
As stated previously, we obtain $\epsilon(q\to 0)= 5.32$ in Kohn-Sham, and we find 4.15 in G$_0$W$_0$-RPA, 3.9 in evQSGW-RPA, and 3.25 in QSGW-RPA, respectively. The GW results are therefore closer to experiment than the Kohn-Sham ones, and the help of the model in avoiding the calculation of the commutators is particularly welcome.

The quality of BSE results obtained using the model screening is shown in the right panel of Fig. \ref{fig:model_vs_abinit}. This picture has been obtained with 864 shifted $\kv$ points,  for both the G$_0$W$_0$+BSE spectra calculated with the model and the full \textit{ab initio} screening. When the model is used in our improved procedure, the differences are very small: we can conclude that we can safely use the model screening in order to converge the BSE results.

\section{Excitons}
\label{sec:excitons}

We can now  analyze the influence of the excitonic effects on the optical spectra of AgCl. 
We will make a detailed analysis of the character of the excitons, in order to understand their spatial localization and interpret the optical properties of AgCl.

\subsection{Excitonic effects: role of the screening and comparison to experiment}

\begin{figure}
	\centering
	\includegraphics[width=\columnwidth]{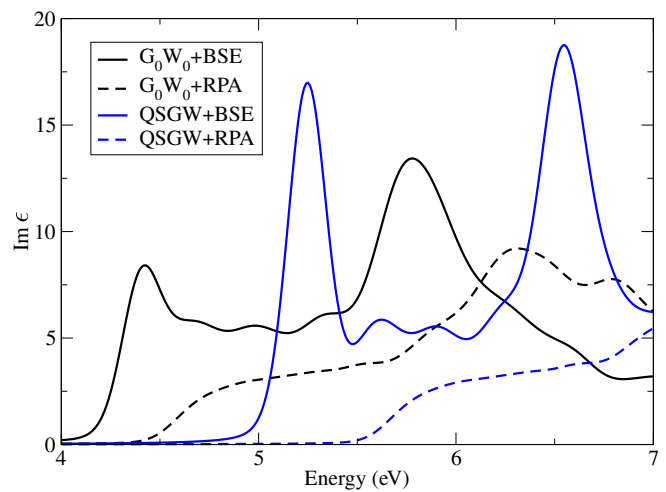}
	\caption{Absorption spectra calculated in the RPA (dashed lines) and from the BSE (solid lines), using scissor corrections and model screenings obtained from G$_0$W$_0$ (black lines) or QSGW (blue lines).}
	\label{GW_scGW}
\end{figure}

Fig. \ref{GW_scGW} shows the absorption spectra obtained, with a converged  grid of 6912 $\bfk$ points, comparing different flavors or RPA and BSE results.  All spectra are calculated with LDA wavefunctions. G$_0$W$_0$+BSE and QSGW+BSE spectra differ for two reasons.
First, a different scissor correction: 1.9 eV or 3.0 eV to simulate the  G$_0$W$_0$ or QSGW bandgap opening, respectively;
Second, the model screening Eq. \eqref{eq:model-eps} is evaluated with  $\epsilon_0= 5.32$ and $\epsilon_0= 3.25$ for the G$_0$W$_0$+BSE and QSGW+BSE spectra, respectively.
The energy of the first exciton peak (solid lines) moves from 4.4 eV in G$_0$W$_0$+BSE to 5.25 eV in QSGW+BSE.
By comparing the BSE spectra with the corresponding RPA results (dashed lines), we find that the reduced screening in
QSGW+BSE crucially enhances excitonic effects with respect to G$_0$W$_0$+BSE. 
The exciton binding energy for the first peak is 0.21 eV within G$_0$W$_0$+BSE and becomes 0.43 eV within QSGW+BSE. 
The larger redshift of the QSGW+BSE spectra partially compensates the larger  scissor correction in the QSGW-RPA result with respect to G$_0$W$_0$-RPA.
Most importantly, the oscillator strength of the excitonic peak is greatly increased.

\begin{figure}
	\centering
\includegraphics[width=\columnwidth]{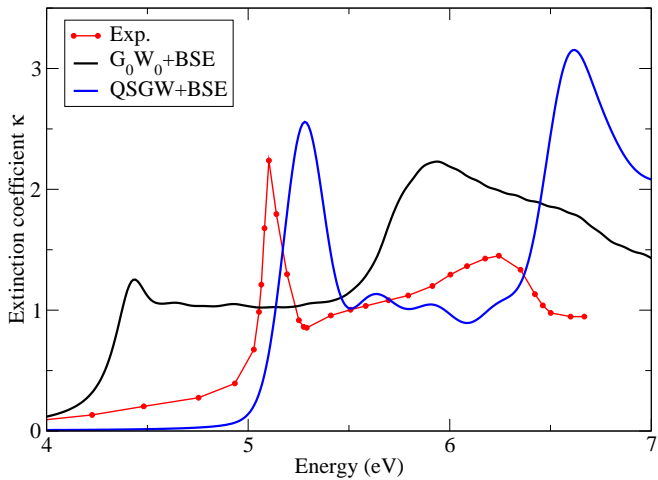}
\caption{Extinction coefficient spectra: comparison between experimental results\cite{Carrera1971}, and G$_0$W$_0$+BSE and QSGW+BSE calculations. }
\label{exp_vs_scGW}
\end{figure}

Our final results for the extinction coefficient are shown in  Fig. \ref{exp_vs_scGW}. The combined use of the converged $\bfk$-point grid and QSGW ingredients, which was made possible thanks to the model screening, improves remarkably the comparison with experiment (red line) with respect to the G$_0$W$_0$+BSE spectra (black line) in Fig. \ref{exp_vs_bse}.
The QSGW+BSE results (blue line) reproduce the first sharp excitonic peak and place it very close to the experimental peak at 5.1 eV.

\subsection{Analysis of the exciton}

Thanks to the full diagonalisation of the  BSE hamiltonian, which provides the eigenvalues $\bar E_\lambda$ and eigenvectors  $\bar A_\lambda$ (see Sec. \ref{subsec:bse}), we can now  analyse in detail the character of the lowest-energy excitons.

The sharp peak in the spectrum is due to three degenerate exciton states.
They are not the lowest-energy excitons though: approximately 50 meV below them there is also a twofold degenerate exciton that does not contribute to the absorption spectrum, i.e. it is dark.

These five excitons originate from transitions between the top-valence and the bottom-conduction bands at the $\bfk$ points close to the minimum direct gaps in the band structure (see Sec. \ref{sec:bandstructure}).  In this region, around the $\Gamma$ point, the valence and conduction bands are almost parallel, which means that many transitions of  similar energy can mix and favour strong excitonic effects \cite{Ehrenreich1963,Harrison1966}.

One notable difference between the dark and bright excitons is the coefficient $\bar A_\lambda$ at $\kv=\Gamma$, which contributes only for the bright excitons. 
Moreover, for the dark excitons the coefficients at $\kv\neq0$ are even functions of $\kv$: $\bar A_\lambda(\kv) = \bar A_\lambda(-\kv)$, while for the bright excitons they are odd: $\bar A_\lambda(\kv) = -\bar A_\lambda(-\kv)$. Since for the same transitions the oscillator strengths are also odd: $\tilde \rho(\kv) = -\tilde \rho(-\kv)$,
the products $\bar A_\lambda(\kv) \tilde \rho(\kv)$ interfere constructively for the bright excitons and destructively for the dark excitons. This different mixing of formerly independent transitions is a typical manifestation of the many-body excitonic effects.

\begin{figure*}[!ht]
    \includegraphics[width=0.95\textwidth]{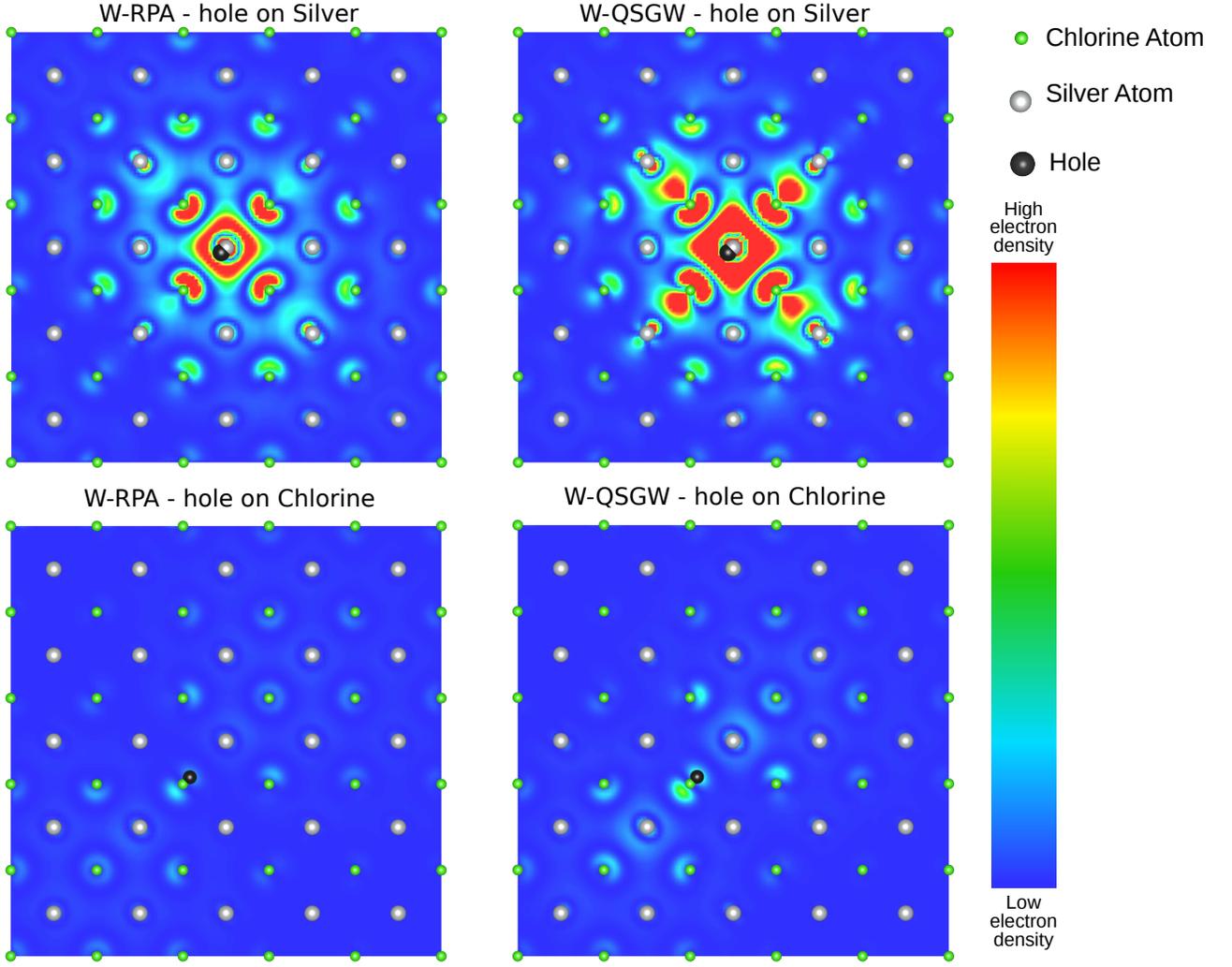}
    \caption{Electron density distribution for the two degeenrate lowest-energy dark excitons.
    Cuts along the [101] plane of for W RPA (left panels) and W QSGW (right panels) exciton in which the hole has been placed close to the Ag atom (top panels) and close to the Cl atom (bottom panels).  Cl (Ag) atoms are represented by  green (grey) balls, while the hole position is black.  \label{fig:excwfs_dark}}
\end{figure*}

\begin{figure*}
    \centering
    \includegraphics[width=0.95\textwidth]{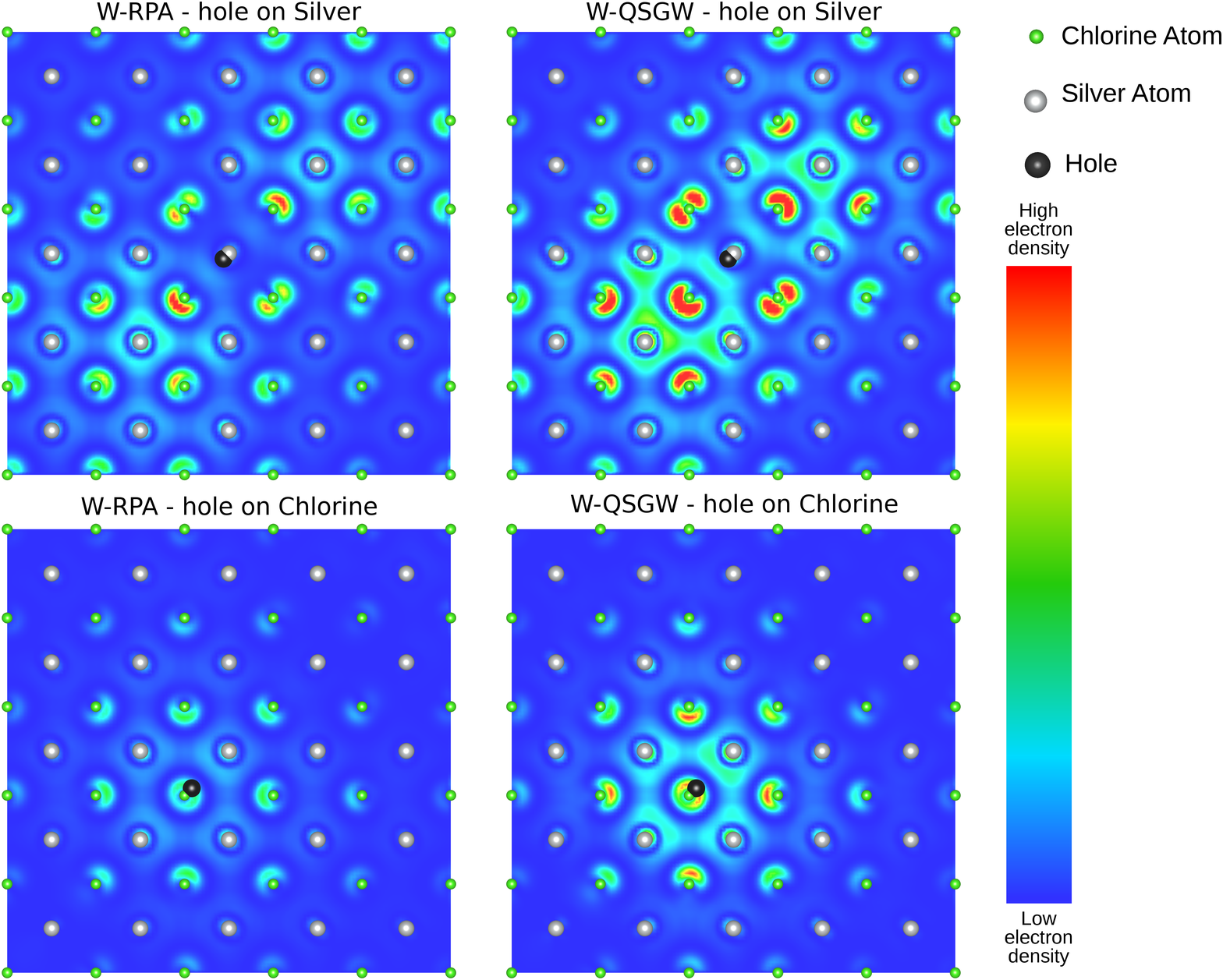}
    \caption{Same as the previous figure for the three degenerate bright excitons. }
    \label{fig:excwfs_bright}
\end{figure*}

We can further contrast the character of the dark and bright excitons by examining their electron-hole correlation function $\Psi_\lambda(\bfr_h,\bfr_e)$.
Fixing the hole at position $\bfr_0$, its square modulus gives the electron probability distribution in the electron-hole pair (in each case we take the sum over the degenerate states $\lambda$).  
 The electron-hole interaction correlates the position of the hole and the electron, it is thus necessary to compare different positions for the hole.
Since the top-valence band has a mixed Ag $4d$ - Cl $3p$ character,
two different locations for the hole are considered: close to the Ag or the Cl sites. In order to avoid the nodes of the valence wave functions, the hole position is slightly shifted away from the atomic sites.
Therefore the electron density plots do not have the cubic symmetry of the crystal.
The corresponding electron distributions are displayed in Figs. \ref{fig:excwfs_dark} and \ref{fig:excwfs_bright} in the color contour plots\footnote{For these plots we have used the VESTA software\cite{Momma2011}.} that represent a cut in the [101] plane of AgCl: the upper panels are for the hole located close to Ag atoms and the bottom panels for the hole close to the Cl atoms; the left panels correspond to calculation using W in the RPA (W-RPA), while the right panels to the calculation using W in the QSGW approach (W-QSGW).
The first figure shows the sum of the electron distribution for the  two degenerate dark excitons and the second one the sum of the three degenerate bright excitons. The saturation intensity is the same in each column but it is 1.8 time larger for the dark than for the bright exciton. 

 The spatial distributions obtained from W-RPA and W-QSGW
 are qualitatively similar. However, the W-QSGW calculation  significantly increases the electron density close to the hole, consistently with the fact that the self-consistent screening leads to a stronger electron-hole interaction. 

The analysis of these plots reveals several aspects of the excitons. 
In both the dark and bright excitons, the intensity is stronger when the hole is placed close to silver than close to chlorine. This difference is more evident for the dark exciton than for the bright one. More importantly, the bright exciton is more delocalised than the dark exciton, whose envelope has a spatial extension that is smaller than 2 unit cells. 
In the bright exciton, when the hole is placed close to a Ag site, there is some intensity around silver atoms, 
but the electron density is  mainly localized around chlorine atoms;
when the hole is instead placed close to a Cl site, the electron density is again mostly localized around chlorine atoms. 
It is interesting here to make a comparison with LiF, which shares the same crystal structure as AgCl. Since LiF is a wide-gap insulator, 
one would expect a tightly bound electron-hole pair with the hole located at F sites and the electron at neighboring Li sites. BSE calculations \cite{Rohlfing1998,Rohlfing2000,Gatti2013} instead have shown that the electron charge is always localised on F atoms (and only weakly on Li atoms).
In other words, the role of Cl in the exciton of AgCl is analogous to  F in LiF.

As can be seen in Fig.\ref{fig:excwfs_dark} and Fig.\ref{fig:excwfs_bright}, the bright exciton with a hole close to chlorine and the dark exciton with hole close to silver have a
spherical shape, whereas the other two cases present an elongated shape. 
An explanation for the difference between these pictures can be drawn from the dominant single-particle transitions that give rise to each exciton. 
We distinguish the two possible hole locations:
If the hole is situated at a silver atom,  
the dark exciton is formed by the dipole-forbidden transition Ag $3d \to 4s$, yielding a spherical shape to the electron distribution;  
instead, the bright exciton, thanks to the Cl-Ag hybridization of the valence band, has the character of a dipole-allowed transition Ag $d \to$ Cl, giving rise to an axial electron distribution.
If the hole is located at a chlorine atom, 
the dark exciton has the character of the dipole-forbidden transition Cl $3p \to$ Cl $3p$ and the corresponding electron distribution has an axial distribution; instead, the character of the bright exciton is the dipole-allowed Cl $3p \to$ Cl $4s$, which is possible thanks to the fact that the Cl $4s$ contribute to  the conduction band (see Fig. ~\ref{fig:dos})  around the $\Gamma$ point where the exciton is formed. This results again into a spherical shape of the electron distribution.

\section{Conclusions}
\label{sec:conclusions}

In conclusion, we have presented an extensive theoretical study of the electronic and optical properties of silver chloride, using \textit{ab initio} calculations starting from KS-DFT. Since the KS band structure severely underestimates all gaps, we have evaluated quasi-particle corrections using the GW approximation of MBPT. We have found that self-consistency is needed to produce realistic results. In order to understand the optical spectra, we had to include the electron-hole interaction, which leads to strong excitonic effects. One could in principle obtain optical spectra also from the computationally more efficient TDDFT, but excitonic  effects in AgCl could not be captured by current approximations. We therefore had to solve the Bethe-Salpeter equation of MBPT. 

Since convergence necessitates a dense Brillouin zone sampling, we used a model to describe the screening of the electron-hole interaction. The most important input for this model is the dielectric constant at vanishing wavevector, which is more difficult to determine than values at larger wavevector. We have shown that this can lead to large errors. We therefore propose to change the model input such that vanishing wavevectors can be avoided. We have shown that this leads to very good agreement between results that are calculated with the full \textit{ab initio} screening, and those obtained using the model screening. As a byproduct, it also allows us to determine dielectric constants using different levels of theory in a very efficient way. 
Using this approach, we have obtained optical spectra in good agreement with experiment, again pointing out the need for self-consistency in the calculations. 

The calculations show that a threefold degenerate bright exciton, which corresponds to the strong peak that is visible at the onset in the experiment, is preceded by a twofold degenerate  exciton that is dark due to destructive interference. Analysis of the electron-hole correlation function reveals that a hole close to a silver atom leads to  a strong redistribution of the electron density, whereas the effect is much weaker for a hole close to a chlorine atom. The use of a self-consistent screening calculated in QSGW strongly enhances the localization of the electron around the hole, with respect to RPA screening of the electron-hole interaction. This is interesting, as it may have consequences for the coupling of electronic excitations in AgCl to the lattice, or for the migration of charge between AgCl as a substrate and molecules adsorbed on its surface.

\acknowledgements

This work was supported by a grant from the Ile-de-France Region – DIM ``Mat\'eriaux anciens et patrimoniaux''.
Computational time was granted by GENCI (Project No. 544).
We acknowledge fruitful discussions with V. de Seauve, M.-A. Languille and B. Lav\'edrine.

\bibliographystyle{apsrev4-1}
\bibliography{ma_biblio}

\end{document}